\documentclass[12pt]{article}
\begin{document}
\newcommand{\beq}{\begin{equation}}
\newcommand{\eeq}{\end{equation}}
\newcommand{\beqa}{\begin{eqnarray}}
\newcommand{\eeqa}{\end{eqnarray}}
\newcommand{\beqar}{\begin{eqnarray*}}
\newcommand{\eeqar}{\end{eqnarray*}}
\newcommand{\al}{\alpha}
\newcommand{\be}{\beta}
\newcommand{\del}{\delta}
\newcommand{\D}{\Delta}
\newcommand{\eps}{\epsilon}
\newcommand{\ga}{\gamma}
\newcommand{\Ga}{\Gamma}
\newcommand{\ka}{\kappa}
\newcommand{\nn}{\nonumber}
\newcommand{\inn}{\!\cdot\!}
\newcommand{\h}{\eta}
\newcommand{\ii}{\iota}
\newcommand{\kk}{\varphi}
\newcommand\F{{}_3F_2}
\newcommand{\la}{\lambda}
\newcommand{\La}{\Lambda}
\newcommand{\na}{\prt}
\newcommand{\Om}{\Omega}
\newcommand{\om}{\omega}
\newcommand\dS{\dot{\cal S}}
\newcommand\dB{\dot{B}}
\newcommand\dG{\dot{G}}
\newcommand\ddG{\dot{\dot{G}}}
\newcommand\dP{\dot{\phi}}
\newcommand{\p}{\phi}
\newcommand{\sig}{\sigma}
\renewcommand{\t}{\theta}
\newcommand{\z}{\zeta}
\newcommand{\ssc}{\scriptscriptstyle}
\newcommand{\eg}{{\it e.g.,}\ }
\newcommand{\ie}{{\it i.e.,}\ }
\newcommand{\labell}[1]{\label{#1}} 
\newcommand{\reef}[1]{(\ref{#1})}
\newcommand\prt{\partial}
\newcommand\veps{\varepsilon}
\newcommand{\pol}{\varepsilon}
\newcommand\vp{\varphi}
\newcommand\ls{\ell_s}
\newcommand\cF{{\cal F}}
\newcommand\cA{{\cal A}}
\newcommand\cS{{\cal S}}
\newcommand\cT{{\cal T}}
\newcommand\cV{{\cal V}}
\newcommand\cL{{\cal L}}
\newcommand\cM{{\cal M}}
\newcommand\cN{{\cal N}}
\newcommand\cG{{\cal G}}
\newcommand\cH{{\cal H}}
\newcommand\cI{{\cal I}}
\newcommand\cJ{{\cal J}}
\newcommand\cl{{\iota}}
\newcommand\cP{{\cal P}}
\newcommand\cQ{{\cal Q}}
\newcommand\cg{{\it g}}
\newcommand\cR{{\cal R}}
\newcommand\cB{{\cal B}}
\newcommand\cO{{\cal O}}
\newcommand\tcO{{\tilde {{\cal O}}}}
\newcommand\bg{\bar{g}}
\newcommand\bb{\bar{b}}
\newcommand\bH{\bar{H}}
\newcommand\bX{\bar{X}}
\newcommand\bK{\bar{K}}
\newcommand\bA{\bar{A}}
\newcommand\bZ{\bar{Z}}
\newcommand\bxi{\bar{\xi}}
\newcommand\bphi{\bar{\phi}}
\newcommand\bpsi{\bar{\psi}}
\newcommand\bprt{\bar{\prt}}
\newcommand\bet{\bar{\eta}}
\newcommand\btau{\bar{\tau}}
\newcommand\bnabla{\bar{\nabla}}
\newcommand\hF{\hat{F}}
\newcommand\hA{\hat{A}}
\newcommand\hT{\hat{T}}
\newcommand\htau{\hat{\tau}}
\newcommand\hD{\hat{D}}
\newcommand\hf{\hat{f}}
\newcommand\hg{\hat{g}}
\newcommand\hp{\hat{\phi}}
\newcommand\hi{\hat{i}}
\newcommand\ha{\hat{a}}
\newcommand\hb{\hat{b}}
\newcommand\hQ{\hat{Q}}
\newcommand\hP{\hat{\Phi}}
\newcommand\hS{\hat{S}}
\newcommand\hX{\hat{X}}
\newcommand\tL{\tilde{\cal L}}
\newcommand\hL{\hat{\cal L}}
\newcommand\tG{{\widetilde G}}
\newcommand\tg{{\widetilde g}}
\newcommand\tphi{{\widetilde \phi}}
\newcommand\tPhi{{\widetilde \Phi}}
\newcommand\td{{\tilde d}}
\newcommand\tk{{\tilde k}}
\newcommand\tf{{\tilde f}}
\newcommand\ta{{\tilde a}}
\newcommand\tb{{\tilde b}}
\newcommand\tc{{\tilde c}}
\newcommand\tR{{\tilde R}}
\newcommand\teta{{\tilde \eta}}
\newcommand\tF{{\widetilde F}}
\newcommand\tK{{\widetilde K}}
\newcommand\tE{{\widetilde E}}
\newcommand\tpsi{{\tilde \psi}}
\newcommand\tX{{\widetilde X}}
\newcommand\tD{{\widetilde D}}
\newcommand\tO{{\widetilde O}}
\newcommand\tS{{\tilde S}}
\newcommand\tB{{\widetilde B}}
\newcommand\tA{{\widetilde A}}
\newcommand\tT{{\widetilde T}}
\newcommand\tC{{\widetilde C}}
\newcommand\tV{{\widetilde V}}
\newcommand\thF{{\widetilde {\hat {F}}}}
\newcommand\Tr{{\rm Tr}}
\newcommand\tr{{\rm tr}}
\newcommand\STr{{\rm STr}}
\newcommand\hR{\hat{R}}
\newcommand\M[2]{M^{#1}{}_{#2}}

\newcommand\bS{\textbf{ S}}
\newcommand\bI{\textbf{ I}}
\newcommand\bJ{\textbf{ J}}

\begin{titlepage}
\begin{center}

\vskip 2 cm
{\LARGE \bf  Effective action of string theory at order $\alpha'$    \\ \vskip 0.75  cm  in the presence of boundary  }\\
\vskip 1.25 cm
   Mohammad R. Garousi\footnote{garousi@um.ac.ir}

\vskip 1 cm
{{\it Department of Physics, Faculty of Science, Ferdowsi University of Mashhad\\}{\it P.O. Box 1436, Mashhad, Iran}\\}
\vskip .1 cm
 \end{center}

\begin{abstract}
Recently, using the assumption that the string theory effective action at the critical dimension  is background independent,  the  classical on-shell effective action of  the bosonic string theory at order $\alpha'$ in a  spacetime manifold without boundary has been reproduced, up to an overall parameter,    by imposing  the $O(1,1)$ symmetry when the background has a  circle. In the presence of the boundary, we consider a background which has boundary and a circle such that the unit normal vector  of the boundary is independent of the circle. Then the   $O(1,1)$ symmetry  can fix  the   bulk action without using the lowest order equation of motion. Moreover, the above constraints and the constraint from the principle of  the least action in the presence of boundary can  fix  the  boundary action, up to five boundary parameters. In the least action principle, we  assume that not only the values of the massless fields but also the values of their first derivatives  are arbitrary on the boundary.

We have also observed that the cosmological/one-dimensional  reduction of the leading order action in the presence of the Hawking-Gibbons boundary term, produces zero  boundary action. Imposing this as another constraint on the boundary couplings at order $\alpha'$, we find the boundary action up to two parameters. For a specific value for these two
parameters, the gravity couplings in the boundary become the Chern-Simons gravity  plus another term which has the Laplacian of the extrinsic curvature.

\end{abstract}
\end{titlepage}

\section{Introduction}

 String theory is a quantum theory of gravity   with  a finite number of massless fields and a  tower of infinite number of  massive fields reflecting the stringy nature of the gravity. The critical dimension for the bosonic string  is  26, and for the type IIA, type IIB, type I and for the heterotic strings is 10. The type IIB superstring theory  on a spacetime manifold with negative cosmological constant which has boundary is conjectured to be dual to a  conformal field theory on the boundary \cite{Maldacena:1997re}. The string theory is usually explored by studying its effective action  which includes the massless fields and their higher derivative terms. For the spacetime manifolds with boundary, the effective action has both bulk and boundary terms, \ie $\bS_{\rm eff}+\prt\!\! \bS_{\rm eff}$. At the leading order of the derivative, the bulk action should  include the Hilbert-Einstein  action at the critical dimension and   the boundary action should include the corresponding Hawking-Gibbons-York boundary term \cite{York:1972sj,Gibbons:1976ue}. These actions and their appropriate  higher derivative extensions should be produced by specific techniques in the string theory.

 The effective actions in the string theory   have a double expansions. The genus-expansion which includes  the  classical tree-level  and a tower of  quantum  loop-level  corrections, and the   stringy-expansion which is an expansion in terms of  higher derivative couplings  at each loop level.
It has been  shown in \cite{Veneziano:1991ek,Meissner:1991zj,Maharana:1992my,Meissner:1996sa} that the tree-level effective action of the bosonic string theory at orders $\alpha'^0$ and $\alpha'$  are invariant under $O(d,d)$ transformations if one compactifies the theory on the  tours $T^d$ and keeps only the zero modes (cosmological reduction).  Using the string field theory, it has been  proved in \cite{Sen:1991zi} that the  cosmological reduction of the tree-level effective action of the bosonic string theory to all orders of $\alpha'$ should be  invariant under $O(d,d)$ transformations. This has been extended in \cite{Hohm:2014sxa} to the classical effective action of the heterotic string theory.

The Einstein theory of general relativity is background independent in the sense that only gauge symmetry is required to specify the theory. We expect that the string theory classical effective action at the critical dimension which is a higher-derivative extension of the Einstein theory at the critical dimension, to be background independent too. Unlike the Einstein action which has only one coupling, however, there are many gauge invariant couplings in the effective action of the string theory at each order of $\alpha'$, \eg at the leading order the independent gauge invariant couplings in the bosonic string theory are
\beqa
\bS_0&=& -\frac{2}{\kappa^2}\int d^{26}x e^{-2\Phi}\sqrt{-G}\,  \left(\alpha_1 R + \alpha_2\nabla_{a}\Phi \nabla^{a}\Phi+\alpha_3 H^2\right)\,.\labell{S0b}
\eeqa
where $\alpha_1,\alpha_2,\alpha_3$ are three parameters. The first term is the Einstein action at the critical dimension 26.
   The background independence assumption then  requires these parameters to be independent of the geometry of the spacetime, \ie if the background has the tours  $T^4$ or $K^3$ the value of the coefficients  $\alpha_1,\alpha_2,\alpha_3$ remains the same. However, the coefficients of the gauge invariant couplings in the reduced action do depend on the geometry of the compact spaces. In other words, if one compactifies the above action on $T^4$, the result  would be the same as the compactification on $K^3$ or any other compact manifolds provided that one takes into account all  the corresponding Kaluza-Klein modes.  However, if one ignores the  Kaluza-Klein modes (dimensional reduction), then the actions in the lower dimension have different symmetries corresponding to the compact spaces, \ie the 22-dimensional action in the case of $T^4$ has symmetry $O(4,4)$ which is different than the symmetry of the 22-dimensional action in the case of $K^3$. This means, if one could fix some how the parameters of the effective action at the critical dimension for a particular geometry  in which the reduced action has a specific symmetry, then that parameters would be  valid for any other geometry. For example, if one considers the background to have a circle, then the dimensional reduction of the action should have the $O(1,1)$ symmetry. This symmetry has been used in \cite{Garousi:2019wgz} to fix the parameters in the above action up to an overall factor, \ie
\beqa
\bS_0&=& -\frac{2\alpha_1}{\kappa^2}\int d^{26}x e^{-2\Phi}\sqrt{-G}\,  \left(  R + 4\nabla_{a}\Phi \nabla^{a}\Phi-\frac{1}{12}H^2\right)\,.\labell{S0bf}
\eeqa
which is the standard effective action of the bosonic string theory for $\alpha_1=1$.  At the higher orders of $\alpha'$, there is the complication   that the effective action has the freedom of the higher-derivative field redefinitions \cite{Metsaev:1987zx}. In these cases, the $O(1,1)$ symmetry may fix the parameters of the independent gauge invariant couplings up to the field redefinitions


When the geometry has  one  circle, the constraints from the $Z_2$-subgroup of the $O(1,1)$ symmetry have been used  in \cite{Garousi:2019wgz,Garousi:2019mca} to find the   effective actions of  the bosonic string theory at four- and six-derivative orders in a minimal scheme, up to an overall factor. Assuming there is such symmetry for the classical effective action of the  type II superstring theories as well, all eight-derivative couplings for NS-NS fields have been found in \cite{Razaghian:2018svg,Garousi:2020mqn,Garousi:2020gio,Garousi:2020lof}, up to an overall factor. The background independent assumption then requires   the resulting couplings to be  valid for any other spacetime, up to the field redefinitions.  In fact, the effective actions found in this way  are fully consistent with the sphere-level S-matrix element of four NS-NS vertex operators and with the results from the sigma model \cite{Garousi:2020gio,Garousi:2020lof}. Moreover, when the geometry has the  tours $T^d$, the cosmological reduction of these effective actions  are also fully consistent with the $O(d,d)$ symmetry \cite{Garousi:2021ikb,Garousi:2021ocs}. Assuming also  the classical world-volume effective actions of the non-pertubative branes in the string theory transform covariantly under  the $Z_2$-transformations when the spacetime geometry has a circle, then many already known and unknown world-volume couplings have been found in \cite{Garousi:2017fbe,Mashhadi:2020mzf}.

In applying the $O(1,1)$ symmetry to find the couplings at orders $\alpha,\alpha'^2,\alpha'^3$ in \cite{Garousi:2019wgz,Garousi:2019mca,Razaghian:2018svg,Garousi:2020mqn,Garousi:2020gio,Garousi:2020lof}, one first needs to  find all ${\it independent}$ gauge invariant couplings in the minimal scheme, \ie the couplings which are not related by various Bianchi identities, by the field redefinitions and by total derivative terms. The number of independent couplings at each order of $\alpha'$ is fixed, however, the structure of the gauge invariant couplings depends on how to use the above freedom to find the independent couplings. The number of independent couplings at orders $\alpha'$, $\alpha'^2$, $\alpha'^3$ in the bosonic theory are $8,\, 60,\, 872$, respectively. These couplings in a specific minimal scheme have been found in   \cite{Metsaev:1987zx,Garousi:2019cdn,Garousi:2020mqn}. Then one should impose the $Z_2$-symmetry on these independent couplings to find their corresponding coefficients at each order of $\alpha'$ up to one parameter. The $Z_2$-transformations or T-duality transformations are the Buscher rules \cite{ Buscher:1987sk,Buscher:1987qj} and some  higher derivative corrections at each order of $\alpha'$ which depends on the minimal scheme that one uses for the gauge invariant couplings at that order  \cite{Garousi:2019wgz}. If one does not use the field redefinitions to write the gauge invariant couplings in the minimal scheme, then the $O(1,1)$ constraint can fix the effective action up to many parameters which can be removed by the field redefinition \cite{Garousi:2019wgz}. These parameters appear also in the corrections to the Buscher rules. By changing these corrections, one can change the scheme of the gauge invariant couplings. However, there would be no scheme for which the T-duality transformations are only the Buscher rules.

When the spacetime geometry has one circle, the $Z_2$-symmetry imposes the following constraint on the bulk effective action:
\beqa
 S_{\rm eff}(\psi)&=&S_{\rm eff}(\psi')\labell{TT0}
 \eeqa
where $S_{\rm eff}$ is  the reduction of   $\!\!\bS_{\rm eff}$, $\psi$ represents the massless fields in the base space and $\psi'$ is its $Z_2$-transformations. There are always some  total derivative terms in the base space \cite{Razaghian:2018svg,Garousi:2020mqn,Garousi:2020gio,Garousi:2020lof} which become zero when spacetime has no boundary.
 However, when the spacetime has boundary, the presence of the total derivative terms dictates that  there must be  some couplings on the boundary as well.

For the spacetime manifold which has boundary, using the background independent assumption, one may consider  a geometry that has a  boundary and  one circle.  Then the $Z_2$-symmetry may fix the couplings in the bulk and boundary actions up to field redefinitions. It has been speculated in \cite{Garousi:2019xlf} that, in the presence of the boundary,  the invariance of the classical effective action under the $Z_2$-transformations should be extended as  follows:
The sum of the bulk and the boundary actions,  \ie $\bS_{\rm eff}+\prt\!\! \bS_{\rm eff}$, should be invariant under the   $Z_2$-transformations, \ie
 \beqa
 S_{\rm eff}(\psi)+\prt S_{\rm eff}(\psi)&=&S_{\rm eff}(\psi')+\prt S_{\rm eff}(\psi')\labell{TT}
 \eeqa
where  $\prt S_{\rm eff}$  is  the reduction of the boundary action $\prt\!\! \bS_{\rm eff}$. There might be some  total derivative terms on the boundary of the base space, however, they become zero  using the Stokes's theorem because the boundary of boundary is zero.  In this paper, we are going to impose the above constraint on the effective action of the bosonic string theory at order $\alpha'$. We consider the  background that its boundary is independent of the circle, \ie the unite normal vector to the boundary is invariant under the T-duality transformations at order $\alpha'$. This particular background constrains the corrections to the Buscher rules. We will see that  for this $Z_2$-transformations,  the $O(1,1)$ constraint is not consistent with  the effective action  in the minimal scheme.  In fact to impose the $O(1,1)$ symmetry for this geometry  one should not use the field redefinitions to reduce the number of gauge invariant couplings at order $\alpha'$.

The  constraint \reef{TT} has been used in \cite{Akou:2020mxx} to find the O-plane effective action   at order $\alpha'^2$ in the presence of the boundary in the type II superstring theories in which the Buscher rules have no correction at orders $\alpha',\alpha'^2$.
  The constraint \reef{TT} has been also used in \cite{Garousi:2019xlf} to find the spacetime effective action of the  bosonic string theory at  order  $\alpha'^0$. This constraint fixes the  bulk actions completely, however, it  fixes the boundary action up to some extra boundary parameters.

When spacetime has boundary, however,  there are further constraints on the  boundary  actions from the principle of the least action.  To be able to extremize the bulk effective actions at each order of $\alpha'$, the  boundary should have specific couplings and the massless fields should have  appropriate  values on the boundary.  For example, the Hawking-Gibbons boundary term along with the arbitrariness of  the metric on the boundary is needed  to be able to extremize the Hilbert-Einstein  action, \ie the Einstein's equations are derived by extremizing the Hilbert-Einstein  action against variations of the spacetime metric $G_{\mu\nu}$ which is arbitrary  on the boundary, \ie $\delta G_{\mu\nu}$ and its tangent derivatives along the boundary are zero. The normal derivative of the metric, however, is not arbitrary on the boundary.     The variation of this term   which is not zero, appears on the boundary when one extremizes the bulk  action. The variation of the Hawking-Gibbons boundary term   cancels this normal derivative term on the boundary. We expect similar constraint for the boundary terms in the effective actions of the string theory. However,  if one assumes only the metric is arbitrary on the boundary,  the standard gravity couplings in the  effective actions of the string theory at orders $\alpha'^2, \alpha'^3$ can not be extremized for any boundary couplings. In fact it has been shown in \cite{Myers:1987yn} that only the gravity couplings in the Euler character   can be extremized. We will show that the $O(1,1)$-constraint on the couplings at order $\alpha'$ produces the bulk gravity couplings which are  the same as  the bulk couplings in the Euler character,  however, it produces the boundary couplings which are consistent with  the Chern-Simons form  as well as some other gravity coupling on the boundary.

Hence, in order to be able to extremize the effective action of the string theory at order $\alpha'^n$, we propose that  not only the massless fields  but also their derivatives  up to order $n$ should be arbitrary on the boundary, \ie  the massless fields are arbitrary on the boundary for the effective action at order $\alpha'^0$,  the massless fields and their first derivatives are arbitrary on the boundary for the effective action at order $\alpha'$, and so on. This may be inspired by the fact that the linear differential equation $\frac{d^{(2n+2)}}{dt^{(2n+2)}}x(t)=0$ has specific solution when the functions $x,\frac{dx}{dt},\cdots, \frac{d^n x}{dt^n}$ are known at the initial and the final times. 

 Imposing the $O(1,1)$-symmetry on the most general gauge invariant couplings at order $\alpha'^0$, one finds that the effective actions of the bosonic string theory are fixed up to one extra parameter in the boundary action \cite{Garousi:2019xlf}, \ie
 \beqa
\bS_0+\prt\!\!\bS_0
&=&-\frac{2\alpha_1}{\kappa^2}\Bigg[ \int d^{D}x \sqrt{-G} e^{-2\Phi} \left(  R + 4\nabla_{\mu}\Phi \nabla^{\mu}\Phi-\frac{1}{12}H^2\right)+2\int d^{D-1}\sigma\sqrt{|g|}  e^{-2\Phi}K\Bigg]\nn\\
&&-\frac{2\alpha_5}{\kappa^2}\int d^{D-1}\sigma e^{-2\Phi}\sqrt{| g|}\,  \left(-\frac{1}{2} K+n^{\mu}\nabla_{\mu}\Phi \right)\labell{bbaction}
\eeqa
where $K$ is trace of the extrinsic curvature and $n^\mu$  is  normal vector to the boundary. It is  outward-pointing (inward-pointing) if the boundary is spacelike (timelike).   The above actions are invariant under the $Z_2$-transformation for arbitrary parameters $\alpha_1,\alpha_5$. The sum of the bulk and boundary terms in the first line are $Z_2$-invariant, and the boundary terms in the second line are also invariant under the $Z_2$-transformations. The standard normalization of the Einstein term fixes $\alpha_1=1$. However, the parameters $a_5$ remains arbitrary. In the supestring theory, there are S-duality as well which constrains the parameter $\alpha_5$ to be zero \cite{Garousi:2019xlf}. In the bosonic string theory, however, there is no such symmetry. One can fix this parameter  by the  principle of the least action as follows: Since the action is at two derivative order, only the massless fields are arbitrary on the boundary. In extremizing the bulk action, the normal derivative of the variation of dilaton which is not zero, does not appear on the boundary, whereas, the variation of the boundary action in the second line above produces such a term. The only way to cancel this term, \ie to be able to extremize the bulk and boundary actions, is to set $\alpha_5=0$. Hence, the constraints from the $Z_2$-symmetry and the least action principle, reproduce  the standard bulk and boundary actions at the leading order of $\alpha'$, \ie
\ie
 \beqa
\bS_0+\prt\!\!\bS_0
=-\frac{2}{\kappa^2}\Bigg[ \int d^{D}x \sqrt{-G} e^{-2\Phi} \left(  R + 4\nabla_{\mu}\Phi \nabla^{\mu}\Phi-\frac{1}{12}H^2\right)+ 2\int d^{D-1}\sigma\sqrt{| g|}  e^{-2\Phi}K\Bigg]\labell{baction}
\eeqa
However, it turns out that if one imposes the $Z_2$-constraint \reef{TT} and the constraint from the least action principle  to  the effective actions of the bosonic string theory at order $\alpha'$, one can not fully fix all parameters in the boundary action.

The cosmological reduction of the classical bulk actions must be invariant under $O(d,d)$ transformations \cite{Sen:1991zi}. We expect the cosmological reduction of the boundary actions to be also invariant under the $O(d,d)$ transformations. In the observation that the cosmological reduction of the leading order bulk action is invariant under the $O(d,d)$ symmetry, one removes a total derivative term which is not invariant under the $O(d,d)$ transformations. We find that the cosmological reduction of the Hawking-Gibbons term is not invariant under the $O(d,d)$ transformations either. However, if one keeps track of the total derivative term and transfers it to the boundary by using the Stokes's theorem, one observes that the cosmological reduction of the boundary action at the leading order  becomes invariant under the $O(d,d)$ transformations. In fact it becomes zero. This motivates us  to  speculate that  the cosmological reduction of the classical boundary actions at any order of $\alpha'$ must be invariant under the $O(d,d)$ transformations, and  may even be zero in a specific scheme in which the cosmological bulk action contains only first time-derivatives as in the leading order, \ie
\beqa
\prt\!\! \bS_{\rm eff}^c&=&0\labell{cosm}
\eeqa
The above discussion is valid not only for the cosmological reduction whose boundary is spacelike, \ie $n^2=-1$, but also for any one-dimensional reduction whose boundary is timelike, \ie $n^2=1$.   In this paper, we would like to impose the $Z_2$-constraint \reef{TT}, the constraint from the least action principle and the above constraint on the one-dimensional  reduction of the boundary actions, to fix the effective actions of the bosonic string theory at order $\alpha'$ when the spacetime has boundary.

The outline of the paper is as follows:  In section 2, we use the Bianchi identities and remove the total derivative terms from the bulk action to the  boundary action  to show that there are 20 independent bulk  and 38 independent boundary gauge invariant couplings at order $\alpha'$, without using the field redefinitions. In section 3, using the background independent assumption, we consider a specific background geometry which has a circle and a boundary that its normal vector is independent of the circle. Then using the fact that the circle reduction of the effective action on this background should have $O(1,1)$ symmetry, we constrain the coefficients of the couplings.  In subsection 3.1,  we show that the T-duality constraint in the bulk fixes the 20 parameters in terms of two parameters. We impose a relation between these two parameters by requiring the effective action to have the standard propagator for the $B$-field.  The resulting bulk  action is exactly the one found by  K.A. Meissner up to one overall factor. The T-duality constraint on the bulk couplings  produces also some total derivative terms in the base space which are transferred to the boundary by using the Stokes's theorem.  In section 3.2, we show that the T-duality constraint on the boundary couplings fixes the 38 boundary parameters in terms of the overall  bulk factor  and in terms of 7 boundary parameters.  In section 4, we impose  the constraint from the principle of the least action. Since the independent bulk couplings have no term with three derivatives, extremizing the bulk action produces no constraint on the bulk parameters. However, extremizing the boundary action, one can fix 2 of the 7  boundary parameters.  In section 5, we study the cosmological/one-dimensional  reduction of the  actions.  We find that the   constraint \reef{cosm} on the boundary action fixes  3 of the 5  boundary parameters. 
In section 6, we briefly discuss our results. 

\section{Gauge invariance constraint at order $\alpha'$}

The effective action of the string theory has a double expansions. One expansion is the genus expansion which includes  the  classical sphere-level and a tower of quantum effects. The other one is the  stringy expansion which is an expansion in terms of higher-derivative couplings. The number of derivatives in each  coupling can be accounted by the order of $\alpha'$. When spacetime has boundary, the sphere-level   effective action $\bS_{\rm eff}+\prt\!\!\bS_{\rm eff}$ has the following   $\alpha'$-expansion in the string frame:
\beqa
\bS_{\rm eff}&=&\sum^\infty_{m=0}\alpha'^m\bS_m=\bS_0+\alpha' \bS_1 +\cdots  ; \qquad\bS_m=-\frac{2}{\kappa^2}\int_M d^D x\sqrt{-G} e^{-2\Phi}\mathcal{L}_m\labell{eq.1}\\
\prt\!\!\bS_{\rm eff}&=&\sum^\infty_{m=0}\alpha'^m\prt\!\!\bS_m=\prt\!\!\bS_0+\alpha' \prt\!\!\bS_1 +\cdots  ;\qquad\prt\!\!\bS_m=-\frac{2}{\kappa^2}\int_{\prt M} d^{D-1} \sigma\sqrt{|g|} e^{-2\Phi}\prt\mathcal{L}_m\nn
\eeqa
where $G$ is determinant of the bulk metric $G_{\mu\nu}$ and boundary is specified by the functions $x^\mu=x^\mu(\sigma^{\tilde{\mu}})$. In the second line, $g$ is determinant of the induced metric on the boundary
\beqa
g_{\tilde{\mu}\tilde{\nu}}&=& \frac{\prt x^\mu}{\prt \sigma^{\tilde{\mu}}}\frac{\prt x^\nu}{\prt \sigma^{\tilde{\nu}}}G_{\mu\nu}\labell{indg}
\eeqa
The effective action must be invariant under the coordinate transformations and under the $B$-field   gauge transformations. On can easily find the independent couplings in the bulk and boundary actions at order $\alpha'^0$, \ie
\beqa
\cL_0&=& \alpha_1 R + \alpha_2\nabla_{\mu}\Phi \nabla^{\mu}\Phi+\alpha_3 H^2\nn\\
\prt\cL_0&=&\alpha_4K+\alpha_5n^{\mu}\nabla_{\mu}\Phi \labell{S0b1}
\eeqa
 where $\alpha_1,\cdots, \alpha_5$ are 5 parameters that the gauge symmetry can not fix them. Using the background independent assumption, they can be fixed by the $Z_2$-symmetry  \reef{bbaction},   and by the least action principle \reef{baction}.

 Using the package   "xAct" \cite{Nutma:2013zea}, one finds there are 41 gauge invariant couplings in the bulk action at order $\alpha'$. However, they are not all independent.   To find the independent bulk couplings, we note that the total derivative terms in the  bulk can be transferred to the boundary using the Stoke's theorem. Hence, the couplings in the bulk should not include total derivative terms.  Moreover, the independent couplings should not be related to each others by the Bianchi identities
\beqa
 R_{\alpha[\beta\gamma\delta]}&=&0\nn\\
 \nabla_{[\mu}R_{\alpha\beta]\gamma\delta}&=&0\labell{bian}\\
\nabla_{[\mu}H_{\alpha\beta\gamma]}&=&0\nn\\
{[}\nabla,\nabla{]}\mathcal{O}-R\mathcal{O}&=&0\nn
\eeqa
Removing the above freedoms from the most general gauge invariant couplings in the bulk action,  one finds there are 20  even-parity independent couplings \cite{Metsaev:1987zx}, \ie
\beqa
\mathcal{L}_1&= &  a_{7}^{} R^2 + a_{6}^{} R H_{\alpha \beta \gamma } H^{\alpha
\beta \gamma } + a_{1}^{} H_{\alpha }{}^{\delta \epsilon }
H^{\alpha \beta \gamma } H_{\beta \delta }{}^{\varepsilon } H_{
\gamma \epsilon \varepsilon } + a_{2}^{} H_{\alpha \beta
}{}^{\delta } H^{\alpha \beta \gamma } H_{\gamma }{}^{\epsilon
\varepsilon } H_{\delta \epsilon \varepsilon }\nn\\&& + a_{3}^{}
H_{\alpha \beta \gamma } H^{\alpha \beta \gamma } H_{\delta
\epsilon \varepsilon } H^{\delta \epsilon \varepsilon } +
a_{4}^{} H_{\alpha }{}^{\gamma \delta } H_{\beta \gamma \delta
} R^{\alpha \beta } + a_{5}^{} R_{\alpha \beta } R^{\alpha
\beta } + a_{8}^{} R_{\alpha \beta \gamma \delta } R^{\alpha
\beta \gamma \delta } \nn\\&&+ a_{9}^{} H_{\alpha }{}^{\delta
\epsilon } H^{\alpha \beta \gamma } R_{\beta \gamma \delta
\epsilon } + a_{11}^{} R \nabla_{\alpha }\nabla^{\alpha }\Phi
+ a_{10}^{} H_{\beta \gamma \delta } H^{\beta \gamma \delta }
\nabla_{\alpha }\nabla^{\alpha }\Phi + a_{13}^{} R
\nabla_{\alpha }\Phi \nabla^{\alpha }\Phi \nn\\&&+ a_{12}^{}
H_{\beta \gamma \delta } H^{\beta \gamma \delta }
\nabla_{\alpha }\Phi \nabla^{\alpha }\Phi + a_{14}^{}
\nabla_{\alpha }\Phi \nabla^{\alpha }\Phi \nabla_{\beta
}\nabla^{\beta }\Phi + a_{15}^{} H_{\alpha }{}^{\gamma \delta
} H_{\beta \gamma \delta } \nabla^{\alpha }\Phi \nabla^{\beta
}\Phi\nn\\&& + a_{16}^{} R_{\alpha \beta } \nabla^{\alpha }\Phi
\nabla^{\beta }\Phi + a_{17}^{} \nabla_{\alpha }\Phi \nabla^{
\alpha }\Phi \nabla_{\beta }\Phi \nabla^{\beta }\Phi +
a_{18}^{} H_{\alpha }{}^{\gamma \delta } H_{\beta \gamma
\delta } \nabla^{\beta }\nabla^{\alpha }\Phi \nn\\&&+ a_{19}^{}
\nabla_{\beta }\nabla_{\alpha }\Phi \nabla^{\beta
}\nabla^{\alpha }\Phi + a_{20}^{} \nabla_{\alpha }H^{\alpha
\beta \gamma } \nabla_{\delta }H_{\beta \gamma }{}^{\delta } \labell{L1bulk}
\eeqa
 where $a_1,\cdots, a_{20}$ are 20 parameters that the gauge symmetry can not fix them. The assumption that the effective action is background independent means  these parameters are background independent.  They may be fixed for the particular geometry which has one circle. Note that the above couplings have no term with three derivatives, hence, in extremizing the above Lagrangian one does not face with the variation of the second derivative of massless fields on the boundary which are non-zero. As a result, our proposal for the  boundary conditions  in which the massless fields and their first derivatives are arbitrary on the boundary, \ie their variations are zero on the boundary,  does not constraint the parameters $a_1,\cdots, a_{20}$. In other words, the above bulk action satisfies $\delta \!\!\bS_1=0$ for any values of the parameters.

Since the boundary of spacetime has a unite normal vector $n^{\mu}$, the boundary Lagrangian  $\prt {\cal L}_1$  should include this vector as well as the  tensors $K_{\mu\nu}$, $H_{\mu\nu\rho}$, $R_{\mu\nu\rho\sigma}$, $\nabla_{\mu}\Phi$ and their derivatives at order $\alpha'$.   The second fundamental form or the extrinsic curvature  of boundary, \ie  $K_{\mu\nu}$, is defined as $ K_{\mu\nu}=P^{\alpha}_{\ \mu}P^{\beta}_{\ \nu}\nabla_{(\alpha}n_{\beta)} $ where
$P^{\mu\nu}=G^{\mu\nu}-n^\alpha n_{\alpha} n^\mu n^\nu$ is the first fundamental form which projects the spacetime tensors tangent to  the boundary.  Using the fact that $n^\mu$ is unit vector orthogonal to the boundary, one can write it as
\beqa
n^{\mu}=\pm (|\prt_\alpha f\prt^\alpha f|)^{-1/2}\prt^\mu f\labell{nf}
\eeqa
where plus (minus)  sign is for timelike (spacelike) boundary in which $n^\mu n_\mu=1$ ($n^\mu n_\mu=-1$), and  the boundary is specified by the function $f$ to be a constant $f^*$. One can rewrite $K_{\mu\nu}$ as
\beqa
K_{\mu\nu}=\nabla_{\mu}n_{\nu}\mp n_{\mu}a_\nu
\eeqa
where minus (plus)  sign is for timelike (spacelike) boundary  and $a_\nu= n^{\rho}\nabla_{\rho}n_{\nu}$ is acceleration. It satisfies the relation $n^\mu a_\mu=0$. Note that the extrinsic curvature is symmetric and satisfies $n^\mu K_{\mu\nu}=0$ and $n^\mu n^\nu\nabla_{\alpha}K_{\mu\nu}=0$ which can easily be seen by writing them in terms of function $f$. Using these symmetries, one finds there are 56 gauge invariant even-parity couplings in the boundary action, \ie
 \beqa
\prt L_1&=& b'_{1} H_{\beta \gamma \delta } H^{\beta \gamma \delta }
K^{\alpha }{}_{\alpha } + b'_{2} H_{\alpha }{}^{\gamma \delta
} H_{\beta \gamma \delta } K^{\alpha \beta } + b'_{3}
K_{\alpha }{}^{\gamma } K^{\alpha \beta } K_{\beta \gamma }
+ b'_{4} K^{\alpha }{}_{\alpha } K_{\beta \gamma } K^{\beta
\gamma }\nn\\&& + b'_{5} K^{\alpha }{}_{\alpha } K^{\beta
}{}_{\beta } K^{\gamma }{}_{\gamma } + b'_{6} H_{\alpha
}{}^{\delta \epsilon } H_{\beta \delta \epsilon } K^{\gamma
}{}_{\gamma } n^{\alpha } n^{\beta } + b'_{7} H_{\alpha
\gamma }{}^{\epsilon } H_{\beta \delta \epsilon } K^{\gamma
\delta } n^{\alpha } n^{\beta } + b'_{8} K^{\alpha \beta }
R_{\alpha \beta } \nn\\&&+ b'_{9} K^{\gamma }{}_{\gamma }
n^{\alpha } n^{\beta } R_{\alpha \beta } + b'_{10}
K^{\alpha }{}_{\alpha } R + b'_{11} K^{\gamma \delta
} n^{\alpha } n^{\beta } R_{\alpha \gamma \beta
\delta } + b'_{12} H^{\beta \gamma \delta } n^{\alpha }
\nabla_{\alpha }H_{\beta \gamma \delta }\nn\\&& + b'_{13} K^{\beta
\gamma } n^{\alpha } \nabla_{\alpha }K_{\beta \gamma } +
b'_{14} K^{\beta }{}_{\beta } n^{\alpha } \nabla_{\alpha
}K^{\gamma }{}_{\gamma } + b'_{15} n^{\alpha }
\nabla_{\alpha }R + b'_{16} H_{\beta \gamma \delta }
H^{\beta \gamma \delta } n^{\alpha } \nabla_{\alpha }\Phi\nn\\&& +
b'_{17} K_{\beta \gamma } K^{\beta \gamma } n^{\alpha }
\nabla_{\alpha }\Phi + b'_{18} K^{\beta }{}_{\beta }
K^{\gamma }{}_{\gamma } n^{\alpha } \nabla_{\alpha }\Phi +
b'_{19} H_{\beta }{}^{\delta \epsilon } H_{\gamma \delta
\epsilon } n^{\alpha } n^{\beta } n^{\gamma } \nabla_{\alpha
}\Phi \nn\\&&+ b'_{20} n^{\alpha } n^{\beta } n^{\gamma }
R_{\beta \gamma } \nabla_{\alpha }\Phi + b'_{21} n^{
\alpha } R \nabla_{\alpha }\Phi + b'_{22} n^{\alpha
} \nabla_{\alpha }\nabla_{\beta }\nabla^{\beta }\Phi +
b'_{23} \nabla_{\alpha }K^{\beta }{}_{\beta } \nabla^{\alpha
}\Phi\nn\\&& + b'_{24} K^{\beta }{}_{\beta } \nabla_{\alpha }\Phi
\nabla^{\alpha }\Phi + b'_{25} \nabla^{\alpha }\Phi \nabla_{
\beta }K_{\alpha }{}^{\beta } + b'_{26} n^{\alpha }
n^{\beta } \nabla_{\alpha }\Phi \nabla_{\beta }K^{\gamma
}{}_{\gamma } + b'_{27} n^{\alpha } \nabla_{\beta
}R_{\alpha }{}^{\beta }\nn\\&& + b'_{28} K^{\gamma
}{}_{\gamma } n^{\alpha } n^{\beta } \nabla_{\alpha }\Phi
\nabla_{\beta }\Phi + b'_{29} \nabla_{\beta }\nabla_{\alpha
}K^{\alpha \beta } + b'_{30} n^{\alpha } n^{\beta }
\nabla_{\beta }\nabla_{\alpha }K^{\gamma }{}_{\gamma } +
b'_{31} K^{\alpha \beta } \nabla_{\beta }\nabla_{\alpha
}\Phi\nn\\&& + b'_{32} K^{\gamma }{}_{\gamma } n^{\alpha }
n^{\beta } \nabla_{\beta }\nabla_{\alpha }\Phi + b'_{33}
\nabla_{\beta }\nabla^{\beta }K^{\alpha }{}_{\alpha } +
b'_{34} K^{\alpha }{}_{\alpha } \nabla_{\beta }\nabla^{\beta
}\Phi + b'_{35} n^{\alpha } \nabla_{\alpha }\Phi
\nabla_{\beta }\nabla^{\beta }\Phi \nn\\&&+ b'_{36} n^{\alpha }
\nabla_{\beta }\nabla^{\beta }\nabla_{\alpha }\Phi + b'_{37}
n^{\alpha } n^{\beta } \nabla_{\beta }\nabla_{\gamma
}K_{\alpha }{}^{\gamma } + b'_{38} H_{\alpha }{}^{\gamma
\delta } H_{\beta \gamma \delta } n^{\alpha } \nabla^{\beta }
\Phi + b'_{39} n^{\alpha } R_{\alpha \beta }
\nabla^{\beta }\Phi \nn\\&&+ b'_{40} K_{\alpha \beta }
\nabla^{\alpha }\Phi \nabla^{\beta }\Phi + b'_{41}
n^{\alpha } \nabla_{\alpha }\Phi \nabla_{\beta }\Phi
\nabla^{\beta }\Phi + b'_{42} n^{\alpha } \nabla_{\beta
}\nabla_{\alpha }\Phi \nabla^{\beta }\Phi \nn\\&&+ b'_{43}
H_{\alpha }{}^{\delta \epsilon } n^{\alpha } n^{\beta }
n^{\gamma } \nabla_{\gamma }H_{\beta \delta \epsilon } +
b'_{44} K^{\beta \gamma } n^{\alpha } \nabla_{\gamma
}K_{\alpha \beta } + b'_{45} K^{\beta }{}_{\beta }
n^{\alpha } \nabla_{\gamma }K_{\alpha }{}^{\gamma } \nn\\&&+
b'_{46} n^{\alpha } n^{\beta } \nabla_{\alpha }\Phi
\nabla_{\gamma }K_{\beta }{}^{\gamma } + b'_{47} n^{\alpha }
n^{\beta } n^{\gamma } \nabla_{\gamma }R_{\alpha
\beta } + b'_{48} n^{\alpha } n^{\beta } n^{\gamma }
\nabla_{\alpha }\Phi \nabla_{\beta }\Phi \nabla_{\gamma
}\Phi \nn\\&&+ b'_{49} n^{\alpha } n^{\beta } \nabla_{\gamma
}\nabla_{\beta }K_{\alpha }{}^{\gamma } + b'_{50} n^{\alpha
} n^{\beta } n^{\gamma } \nabla_{\alpha }\Phi \nabla_{\gamma
}\nabla_{\beta }\Phi + b'_{51} n^{\alpha } n^{\beta }
n^{\gamma } \nabla_{\gamma }\nabla_{\beta }\nabla_{\alpha
}\Phi\nn\\&& + b'_{52} n^{\alpha } n^{\beta } \nabla_{\gamma
}\nabla^{\gamma }K_{\alpha \beta } + b'_{53} n^{\alpha }
n^{\beta } \nabla_{\beta }K_{\alpha \gamma } \nabla^{\gamma
}\Phi + b'_{54} H^{\beta \gamma \delta } n^{\alpha }
\nabla_{\delta }H_{\alpha \beta \gamma }\nn\\&& + b'_{55} H_{\alpha
}{}^{\beta \gamma } n^{\alpha } \nabla_{\delta }H_{\beta
\gamma }{}^{\delta } + b'_{56} n^{\alpha } n^{\beta }
n^{\gamma } n^{\delta } \nabla_{\delta }\nabla_{\gamma
}K_{\alpha \beta }\labell{b0}
 \eeqa
where $b'_1,\cdots b'_{56}$ are 56  parameters. The terms in the boundary action which have bulk fields and have one vector $n^\mu$  can be interpreted as the total derivative terms in the bulk action that are transferred to the boundary by the Stoke's theorem. Note that we have not considered  the curvature tensors and the covariant derivatives that are made of the induced metric \reef{indg}, because they are related to the curvature tensors and covariant derivatives constructed from the spacetime metric by various Gauss-Codazzi  relations.

Not all of the  couplings  in \reef{b0} however are independent. Some of them are related by the Bianchi identities and some others by the total derivative terms in the boundary.  To remove the redundancy corresponding to the total derivative terms, we add to  $\prt L_1$ all total derivative terms at order $\alpha'$ with arbitrary coefficients.
The total derivative terms in the boundary have different structure  than the total derivative terms in the bulk. According to the Stokes's theorem, the total derivative terms in the boundary which have the following structure:
\beqa
\alpha'\int_{\prt M^{(D)} }d^{D-1}\sigma \sqrt{|g |}\cT_1&\equiv&\alpha'\int_{\prt M^{(D)}}d^{D-1}\sigma \sqrt{|g |} n_{\alpha}\nabla_{\beta}(e^{-2\Phi}\mathcal{F}_1^{\alpha\beta})\labell{totb}
\eeqa
are zero because the boundary of boundary is zero (see \eg the appendix in \cite{Akou:2020mxx}). In above equation   $ \mathcal{F}_1^{\alpha\beta} $ is an arbitrary antisymmetric even-parity tensor constructed from $n$, $K$, $\nabla K$, $H^2$, $\nabla\Phi$, $\nabla\nabla\Phi$, $R$ at two-derivative order, \ie
\beqa
\mathcal{F}_1^{\alpha\beta}&=&f_{2} (H^{\beta \delta \epsilon } H_{\gamma \delta \epsilon
} n^{\alpha } n^{\gamma } -  H^{\alpha \delta \epsilon }
H_{\gamma \delta \epsilon } n^{\beta } n^{\gamma }) + f_{3}
(n^{\beta } n^{\gamma } R^{\alpha }{}_{\gamma } -
n^{\alpha } n^{\gamma } R^{\beta }{}_{\gamma }) \nn\\&&+
f_{4} (n^{\gamma } \nabla^{\alpha }K^{\beta }{}_{\gamma } -
 n^{\gamma } \nabla^{\beta }K^{\alpha }{}_{\gamma }) +
f_{5} (n^{\delta } \nabla^{\alpha }K^{\beta }{}_{\delta } -
 n^{\delta } \nabla^{\beta }K^{\alpha }{}_{\delta }) \nn\\&& +
f_{6} (n^{\beta } \nabla^{\alpha }K^{\gamma }{}_{\gamma } -
 n^{\alpha } \nabla^{\beta }K^{\gamma }{}_{\gamma }) +
f_{7} (n^{\beta } \nabla^{\alpha }K^{\delta }{}_{\delta } -
 n^{\alpha } \nabla^{\beta }K^{\delta }{}_{\delta }) \nn\\&& +
f_{8} (K^{\gamma }{}_{\gamma } n^{\beta } \nabla^{\alpha
}\Phi -  K^{\gamma }{}_{\gamma } n^{\alpha } \nabla^{\beta
}\Phi) + f_{9} (K^{\delta }{}_{\delta } n^{\beta } \nabla^{
\alpha }\Phi -  K^{\delta }{}_{\delta } n^{\alpha }
\nabla^{\beta }\Phi) \nn\\&& + f_{10} (n^{\beta } \nabla_{\gamma
}K^{\alpha \gamma } -  n^{\alpha } \nabla_{\gamma }K^{\beta
\gamma }) + f_{11} (n^{\beta } n^{\gamma } \nabla^{\alpha }
\Phi \nabla_{\gamma }\Phi -  n^{\alpha } n^{\gamma }
\nabla^{\beta }\Phi \nabla_{\gamma }\Phi) \nn\\&& + f_{12}
(n^{\beta } n^{\gamma } \nabla_{\gamma }\nabla^{\alpha }\Phi
-  n^{\alpha } n^{\gamma } \nabla_{\gamma }\nabla^{\beta
}\Phi) + f_{13} (K^{\beta }{}_{\gamma } n^{\alpha }
\nabla^{\gamma }\Phi -  K^{\alpha }{}_{\gamma } n^{\beta }
\nabla^{\gamma }\Phi) \nn\\&& + f_{14} (n^{\beta } n^{\gamma } n^{
\delta } \nabla_{\delta }K^{\alpha }{}_{\gamma } -  n^{\alpha
} n^{\gamma } n^{\delta } \nabla_{\delta }K^{\beta
}{}_{\gamma }) + f_{15} (n^{\beta } \nabla_{\delta
}K^{\alpha \delta } -  n^{\alpha } \nabla_{\delta }K^{\beta
\delta })  \nn\\&&+ f_{16} (K^{\beta }{}_{\delta } n^{\alpha }
\nabla^{\delta }\Phi -  K^{\alpha }{}_{\delta } n^{\beta }
\nabla^{\delta }\Phi)
\eeqa
where $f_2,\cdots f_{16}$ are 15 arbitrary parameters. Adding the above total derivative terms   to $\prt L_1$, one finds the same Lagrangian     but with different parameters $b_1, b_2, \cdots$. We call the new Lagrangian  $\prt {\cal L}_1$. Hence
\beqa
{\bf \Delta}-\cT_1&=&0\labell{DLb}
\eeqa
where ${\bf\Delta}=\prt{\cal L}_1-\prt L_1$ is the same as $\prt L_1$ but with coefficients $\delta b_1,\delta b_2,\cdots$ where $\delta b_i= b_i-b'_i$. Solving the above equation, after imposing the Bianchi identities  \reef{bian} and the identities corresponding to  the unit vector $n^{\mu}$, one finds some linear  relations between  only $\delta b_1,\delta b_2,\cdots$ which indicate how the couplings are related among themselves by the total derivative terms,  by the Bianchi identities, and by  the identities corresponding to  the unit vector. The above equation also gives some relations between the coefficients of the total derivative terms and $\delta b_1,\delta b_2,\cdots$ in which we are not interested.

To impose in \reef{DLb} the Bianchi identities \reef{bian}  and the identities corresponding to the unit vector $n$, we write the covariant derivatives and the curvatures in terms of partial derivatives of metric, dilaton, $H$ and $n$. We then write the partial derivatives of $H$ in terms of potential $B$-field and the partial derivatives of $n$ in terms of function $f$, using the definition \reef{nf}. Then all the Bianchi identities \reef{bian} and the identities corresponding to the unit vector $n$ satisfy automatically. In other words, there is no identities any more when one rewrites everything in terms of metric, dilaton, the potential $B$-field and $f$. To simplify  the calculation, one may go to the local frame \cite{Garousi:2019cdn} in which the first partial derivative of metric is zero. All these steps can be done easily by the computer.  Then one finds 38 relations between  only $\delta b_1,\delta b_2,\cdots$ which indicate there are 38 independent couplings.   One particular choice for the independent  boundary couplings is the following:
\beqa
\prt \cL_1&=&b_{1}^{} H_{\beta \gamma \delta } H^{\beta \gamma \delta }
K^{\alpha }{}_{\alpha } + b_{2}^{} H_{\alpha }{}^{\gamma \delta
} H_{\beta \gamma \delta } K^{\alpha \beta } + b_{3}^{}
K_{\alpha }{}^{\gamma } K^{\alpha \beta } K_{\beta \gamma }
+ b_{4}^{} K^{\alpha }{}_{\alpha } K_{\beta \gamma } K^{\beta
\gamma } \nn\\&&+ b_{5}^{} K^{\alpha }{}_{\alpha } K^{\beta
}{}_{\beta } K^{\gamma }{}_{\gamma } + b_{6}^{} H_{\alpha
}{}^{\delta \epsilon } H_{\beta \delta \epsilon } K^{\gamma
}{}_{\gamma } n^{\alpha } n^{\beta } + b_{7}^{} H_{\alpha
\gamma }{}^{\epsilon } H_{\beta \delta \epsilon } K^{\gamma
\delta } n^{\alpha } n^{\beta } + b_{8}^{} K^{\alpha \beta }
R_{\alpha \beta } \nn\\&&+ b_{9}^{} K^{\gamma }{}_{\gamma }
n^{\alpha } n^{\beta } R_{\alpha \beta } + b_{10}^{}
K^{\alpha }{}_{\alpha } R + b_{11}^{} K^{\gamma \delta
} n^{\alpha } n^{\beta } R_{\alpha \gamma \beta
\delta } + b_{12}^{} H^{\beta \gamma \delta } n^{\alpha }
 \nabla_{\alpha }H_{\beta \gamma \delta }\nn\\&& + b_{13}^{} K^{\beta
\gamma } n^{\alpha } \nabla_{\alpha }K_{\beta \gamma } +
b_{14}^{} K^{\beta }{}_{\beta } n^{\alpha } \nabla_{\alpha
}K^{\gamma }{}_{\gamma } + b_{15}^{} n^{\alpha }
 \nabla_{\alpha }R + b_{16}^{} H_{\beta \gamma \delta }
H^{\beta \gamma \delta } n^{\alpha } \nabla_{\alpha }\Phi \nn\\&&+
b_{17}^{} K_{\beta \gamma } K^{\beta \gamma } n^{\alpha }
 \nabla_{\alpha }\Phi + b_{18}^{} K^{\beta }{}_{\beta }
K^{\gamma }{}_{\gamma } n^{\alpha } \nabla_{\alpha }\Phi +
b_{19}^{} H_{\beta }{}^{\delta \epsilon } H_{\gamma \delta
\epsilon } n^{\alpha } n^{\beta } n^{\gamma } \nabla_{\alpha
}\Phi \nn\\&&+ b_{20}^{} n^{\alpha } n^{\beta } n^{\gamma }
R_{\beta \gamma } \nabla_{\alpha }\Phi + b_{21}^{} n^{
\alpha } R \nabla_{\alpha }\Phi + b_{22}^{} K^{\beta
}{}_{\beta } \nabla_{\alpha }\Phi \nabla^{\alpha }\Phi +
b_{23}^{} n^{\alpha } n^{\beta } \nabla_{\alpha }\Phi
 \nabla_{\beta }K^{\gamma }{}_{\gamma }\nn\\&& + b_{24}^{} K^{\gamma
}{}_{\gamma } n^{\alpha } n^{\beta } \nabla_{\alpha }\Phi
 \nabla_{\beta }\Phi + b_{25}^{} n^{\alpha } n^{\beta }
 \nabla_{\beta } \nabla_{\alpha }K^{\gamma }{}_{\gamma } +
b_{26}^{} K^{\alpha \beta } \nabla_{\beta } \nabla_{\alpha
}\Phi\nn\\&& + b_{27}^{} K^{\gamma }{}_{\gamma } n^{\alpha }
n^{\beta } \nabla_{\beta } \nabla_{\alpha }\Phi + b_{28}^{} H_{
\alpha }{}^{\gamma \delta } H_{\beta \gamma \delta }
n^{\alpha } \nabla^{\beta }\Phi + b_{29}^{} n^{\alpha }
R_{\alpha \beta } \nabla^{\beta }\Phi + b_{30}^{}
K_{\alpha \beta } \nabla^{\alpha }\Phi \nabla^{\beta }\Phi \nn\\&&+
b_{31}^{} n^{\alpha } \nabla_{\alpha }\Phi \nabla_{\beta
}\Phi \nabla^{\beta }\Phi + b_{32}^{} n^{\alpha }
 \nabla_{\beta } \nabla_{\alpha }\Phi \nabla^{\beta }\Phi +
b_{33}^{} H_{\alpha }{}^{\delta \epsilon } n^{\alpha }
n^{\beta } n^{\gamma } \nabla_{\gamma }H_{\beta \delta
\epsilon } \nn\\&&+ b_{34}^{} n^{\alpha } n^{\beta } n^{\gamma }
 \nabla_{\alpha }\Phi \nabla_{\beta }\Phi \nabla_{\gamma
}\Phi + b_{35}^{} n^{\alpha } n^{\beta } n^{\gamma } \nabla_{
\alpha }\Phi \nabla_{\gamma } \nabla_{\beta }\Phi + b_{36}^{}
n^{\alpha } n^{\beta } n^{\gamma } \nabla_{\gamma
} \nabla_{\beta } \nabla_{\alpha }\Phi \nn\\&&+ b_{37}^{} n^{\alpha }
n^{\beta } \nabla_{\beta }K_{\alpha \gamma } \nabla^{\gamma
}\Phi + b_{38}^{} n^{\alpha } n^{\beta } n^{\gamma }
n^{\delta } \nabla_{\delta } \nabla_{\gamma }K_{\alpha \beta
}\labell{L1boundary}
\eeqa
 where $b_1,\cdots, b_{38}$ are background independent  boundary parameters. They may however dependent on the type of boundary, \ie timelike or spacelike  boundary. These parameters   may be fixed by imposing the $Z_2$-symmetry, by the least action principle and by the cosmological constraint \reef{cosm}.  Note that the above boundary couplings do have  terms with two and three derivatives, hence, in extremizing the above Lagrangian one  faces with the variation of the second and third derivatives of the massless fields on the boundary which are non-zero. As a result, the least action principle   constrains the parameters $b_1,\cdots, b_{38}$. In other words, the above boundary action satisfies $\delta (\prt\!\!\bS_1)=0$ for some specific  values of the parameters which, as we will see, are consistent with the T-duality.  Note that, as we will see,  if one uses the boundary condition that only the massless fields are arbitrary on the boundary, \ie only the variation of massless fields on the boundary are zero, then  the least action principle would constrain more strongly the parameters $b_1,\cdots, b_{38}$ which would  not be consistent with the T-duality. In the next section we first find the relations between the bulk parameters and the relations between the  boundary and bulk parameters by imposing the $Z_2$-symmetry, and in the  section after we impose the  least action principle   to further constrain the parameters in the boundary action.

\section{$Z_2$-invariance constraint}

Using the assumption that the effective actions in the string theory are independent of the geometry of the spacetime, we now explicitly impose the $Z_2$-symmetry on the effective actions to find some relations between  the parameters in the   gauge invariant couplings \reef{L1bulk} and \reef{L1boundary}.  To this end,  we consider a particular geometry that its bulk and  boundary  have the structures $M^{(D)}=M^{(D-1)}\times S^{(1)}$ and $\prt M^{(D)}=\prt M^{(D-1)}\times S^{(1)}$. The manifold $M^{(D)}$ has coordinates $x^\mu=(x^a,y)$ and  its boundary  $\prt M^{(D)}$ has coordinates $\sigma^{\tilde{\mu}}=(\sigma^\ta, y)$ where $y$ is the coordinate of the circle $S^{(1)}$. The Kaluza-Klein reduction of the metric and the reduction of B-field and dilaton are \cite{Maharana:1992my}
 \beqa
G_{\mu\nu}=\left(\matrix{\bg_{ab}+e^{\varphi}g_{a }g_{b }& e^{\varphi}g_{a }&\cr e^{\varphi}g_{b }&e^{\varphi}&}\!\!\!\!\!\!\right),\, B_{\mu\nu}= \left(\matrix{\bb_{ab}+\frac{1}{2}b_{a }g_{b }- \frac{1}{2}b_{b }g_{a }&b_{a }\cr - b_{b }&0&}\!\!\!\!\!\!\right),\,  \Phi=\bar{\phi}+\varphi/4\labell{reduc}\eeqa
where $\bg_{ab}$ is the metric, $ \bb_{ab}$  is  the   B-field, $\bphi$ is dilaton  and $g_{a},\, b_{b}$ are two vectors  in the base space. The reduction of the unit vector $n^{\mu}$ is
\beqa
n^{\mu}=\left(\matrix{n^a&\cr 0&}\!\!\!\!\!\!\right)\labell{Rn}
\eeqa
where $n^a$ is the unit vector to the boundary in the base space. Using these reductions, then one  reduces the effective actions $\bS_{\rm eff}+\prt\!\!\bS_{\rm eff}$ on the circle to find  $S_{\rm eff}(\psi)+\prt S_{\rm eff}(\psi)$ where $\psi$ represents all the  massless fields in the base space. The $Z_2$-symmetry then constraints the effective action to  satisfy the relation \reef{TT}.
The $Z_2$-transformation of the base space fields in \reef{TT} are  the Buscher rules \cite{Buscher:1987sk,Buscher:1987qj} and their higher derivative corrections, \ie
\beqa
\psi'&=&\sum_{m=0}^\infty\alpha'^m\psi'_m\labell{epsi}
\eeqa
where $\psi'_0$ represents the  Buscher rules. In terms of the reductions \reef{reduc}, they are
\beqa
\varphi'= -\varphi
\,\,\,,\,\,g'_{a }= b_{a }\,\,\,,\,\, b'_{a }= g{_a }\labell{T2}
\eeqa
The base space  metric, dilaton, $\bb$-field and the unite vector $n^a$ are invariant. The  $\psi'_1$ in \reef{epsi} represents two derivative corrections to the Buscher rules, and so on.

The corrections to the Buscher rules depend on the scheme that one uses for the gauge invariant  couplings \cite{Garousi:2019wgz}, and vis versa.   Since we have not used the field redefinitions to write the gauge invariant couplings, we are free to consider a specific geometry for imposing the $O(1,1)$ symmetry. For the geometry that we have considered, the unit normal vector $n^a$ on the boundary is independent of the circle on which the T-duality is  imposed, \ie $\bg_{ab}n^an^b=\pm 1$ is invariant under the T-duality. Hence,  the base space metric must be invariant under the T-duality at any order of $\alpha'$.  On the other hand, under the reduction \reef{reduc}, the density $e^{-2\Phi}\sqrt{-G}$ reduces to $e^{-2\bphi}\sqrt{-\bg}$ which may be invariant, in a specific scheme,  under the T-duality at any order of $\alpha'$. Therefore,  in a specific scheme, the base space dilaton may be invariant at any order of $\alpha'$. 
So, for the background that we consider, the base space unit normal vector $n^a$, the dilaton $\bphi$ and the metric $\bg_{ab}$ do not appear in the T-duality transformations. The base space $\bb$-field which is invariant under the Buscher rules, however, appears in the higher derivative corrections because this field and  the vectors $g_a,\, b_a$ should satisfy a Bianchi identity  \cite{Kaloper:1997ux}. Hence, we consider the T-duality transformations that  involve only  the derivatives of the base space fields $\varphi, g_a, b_a,\bb_{ab}$.

In order the T-duality constraint \reef{TT} to be satisfied, one should  add  some total derivative terms at the boundary which are zero by the Stokes's theorem, \ie
\beqa
T&\equiv&-\frac{2}{\kappa^2}\int_{\prt M^{(D-1)}}d^{D-2}\sigma \sqrt{|\tg |} n_{a}\nabla_{b}(e^{-2\bphi}F^{ab}) =0
\labell{bstokes}
\eeqa
where   $F^{ab}(\psi)$ is an arbitrary antisymmetric tensor constructed from the gauge invariant base space fields. It can be $\alpha'$-expanded as $F^{ab}=\sum_{m=0}^\infty\alpha'^mF_m^{ab}$, which produces an $\alpha'$-expansion for $T= \sum_{m=0}^\infty\alpha'^mT_m$. Then one can study the $Z_2$-constraint \reef{TT} at each order of $\alpha'$.

Replacing  the expansions \reef{eq.1} and \reef{epsi} into the $Z_2$-symmetry \reef{TT}, one finds the following relation at  order $\alpha'^0$:
\beqa
S_0(\psi)+\prt S_0(\psi)&=&S_0(\psi'_0)+\prt S_0(\psi'_0)\labell{S0}
\eeqa
Note that it is impossible to construct the antisymmetric tensor  $F_0^{ab}$ at zero derivative order. Hence there is no total derivative on the boundary in this case. This constraint has been used in \cite{Garousi:2019xlf} to find the effective action \reef{bbaction} at order  $\alpha'^0$.

Using the relation  \reef{S0}, one finds that the constraint \reef{TT} produces the following relation  at order $\alpha'$:
\beqa
S_1(\psi)+\prt S_1(\psi)+T_1(\psi)&=&S_1(\psi'_0)+\prt S_1(\psi'_0)+\Delta S_0+\Delta\prt S_0\labell{S1}
\eeqa
where $T_1(\psi)$ is the total derivative terms \reef{bstokes} at order $\alpha'$, and $\Delta S_0$, $\Delta\prt S_0$ are defined in the following $\alpha'$-expansions:
\beqa
S_0(\psi'_0+\alpha'\psi'_1)-S_0(\psi'_0)&=&\alpha'\Delta S_0+\cdots\nn\\
\prt S_0(\psi'_0+\alpha'\psi'_1)-\prt S_0(\psi'_0)&=&\alpha'\Delta \prt S_0+\cdots\labell{DS0}
\eeqa
where dots represent some terms at higher orders of $\alpha'$ in which we are not interested in this paper. The constraint \reef{S1} can be written as
\beqa
S_1(\psi)-S_1(\psi'_0)-\Delta S_0&=&-T_1(\psi)-\Big(\prt S_1(\psi)-\prt S_1(\psi'_0)-\Delta\prt S_0\Big)\labell{S11}
\eeqa
The terms on the left-hand side are in the $(D-1)$-dimensional base space whereas the terms on the right-hand side are in its boundary. Only the total derivative terms in the  $(D-1)$-dimensional action can contribute to its boundary  action.
Hence,  the bulk actions on the left-hand side should be  some total derivative terms, \ie
\beqa
 S_1(\psi)-S_1(\psi'_0)-\Delta S_0&=&\frac{2}{\kappa^2}\int_{M^{(D-1)}} d^{D-1}x\sqrt{-\bg}\nabla_a (A_1^a e^{-2\bphi})\labell{S1b}
 \eeqa
where  $A_1^a$ is a vector made of the covariant derivative of the massless fields in the base space at order $\alpha'$. 
The total derivative terms then  produce some boundary terms using the Stokes's theorem.

The Stokes's theorem in the base space is
\beqa
\int_{M^{(D-1)}} d^{D-1}x\sqrt{-\bg}\nabla_a (A_1^a e^{-2\bphi})&=&\int_{\prt M^{(D-1)}}d^{D-2}\sigma\sqrt{|\tg|}n_{a} A_1^a e^{-2\bphi}\labell{stok}
\eeqa
where   $n^a$ is the unit  vector orthogonal  to the boundary in the base space and the boundary is specified by the  functions  $x^a=x^a(\sigma^\ta)$.  The unit vector is outward-going (inward-going) if the boundary is spacelike (timelike). The metric in the square root on the right-hand side is    the induced metric, \ie
\beqa
\tg_{\ta\tb}&=&\frac{\prt x^{a}}{\prt \sigma^{\ta}}\frac{\prt x^{b}}{\prt \sigma^{\tb}}\bg_{ab}\labell{gtatb}
 \eeqa
Using the Stokes's theorem to transfer  the bulk total derivative terms on the right-hand side of \reef{S1b} to the boundary, then the $Z_2$-symmetry on the effective action at order $\alpha'$ produces the bulk constraint \reef{S1b} as well as the following constraint on the boundary couplings:
\beqa
\prt S_1(\psi)-\prt S_1(\psi'_0)-\Delta\prt S_0+T_1(\psi)+\frac{2}{\kappa^2}\int_{\prt M}d^{D-2}\sigma\sqrt{|\tg|}n_{a} A_1^a e^{-2\bphi}&=&0\labell{S11b}
\eeqa
where $A_1^a$ has to be found from the bulk constraint \reef{S1b}.  The constraint \reef{S1b} and \reef{S11b}  produce some relations between the  parameters in the gauge invariant couplings \reef{L1bulk} and \reef{L1boundary}.

\subsection{Bulk constraint}

The bulk constraint \reef{S1b} has been used in \cite{Garousi:2019wgz} to find  some relations between the parameters in the  action  \reef{L1bulk}  for the most general T-duality transformation at order $\alpha'$.  The total derivative terms have been ignored in calculating $\Delta S_0$ in \cite{Garousi:2019wgz}
 because in that calculation it was assumed spacetime has  no boundary. Hence, the result in \cite{Garousi:2019wgz} can not be used for the present case that there is boundary. In the present case all  total derivative terms in the base space should be taken into account.  So we solve the constraint \reef{S1b} in this section to keep track of the total derivative terms carefully.

To find $\Delta S_0$ from the expansion \reef{DS0}, one needs the reduction of the bulk action at order $\alpha'^0$. Using the reductions \reef{reduc}, one finds the reduction of the bulk action $\!\!\bS_0$ to be
\beqa
S_0(\psi)&=& -\frac{2}{\kappa^2}\int d^{D-1}x e^{-2\bphi}\sqrt{-\bg}\Big[\bar{R}-\nabla^a\nabla_a\vp-\frac{1}{4}\nabla_a\vp \nabla^a\vp-\frac{1}{4}(e^{\vp}V^2 +e^{-\vp}W^2)\nn\\
&&\qquad\qquad\qquad\qquad\qquad+4\nabla_a\bphi\nabla^a \bphi+2\nabla_a\bphi\nabla^a\vp-\frac{1}{12}\bH_{abc}\bH^{abc}\Big]\labell{S0r}
\eeqa
 where $V_{ab}$ is field strength of the $U(1)$ gauge field $g_{a}$, \ie $V_{ab}=\prt_{a}g_{b}-\prt_{b}g_{a}$, and $W_{\mu\nu}$ is field strength of the $U(1)$ gauge field $b_{a}$, \ie $W_{ab}=\prt_{a}b_{\nu}-\prt_{b}b_{a}$. The    three-form $\bH$ is defined as $\bH_{abc}=\hat{H}_{abc}-\frac{3}{2}g_{[a}W_{bc]}-\frac{3}{2}b_{[a}V_{bc]}$ where the three-form  $\hat{H}$ is field strength of the two-form $\bb_{ab}  $ in \reef{reduc}.   Since $\bH$ is not exterior derivative of a two-form,  it satisfies  anomalous Bianchi identity, whereas the $W,V$ satisfy the ordinary Bianchi identity, \ie
 \beqa
 \prt_{[a} \bH_{bcd]}&=&-\frac{3}{2}V_{[ab}W_{cd]}\labell{anB}\\
 \prt_{[a} W_{bc]}&=&0\nn\\
  \prt_{[a} V_{bc]}&=&0\nn
 \eeqa
  Our notation for making  antisymmetry  is such that \eg $g_{[a}W_{bc]}=\frac{1}{3}(g_aW_{bc}-g_{b}W_{ac}-g_cW_{ba})$.

The $Z_2$-transformations at order $\alpha'^0$, \ie $\psi_0'$, are given by the Buscher rules \reef{T2}, and at order $\alpha'$ are given by  the following:
\beqa
\varphi'= -\varphi+\alpha'\Delta\vp,\,\,g'_{a }= b_{a }+\alpha'e^{\vp/2}\Delta g _a,\,\, b'_{a }= g_{a }+\alpha'e^{-\vp/2}\Delta b _a,\,\,\bH_{abc}'=\bH_{abc}+\alpha'\Delta\bH_{abc} \labell{T22}
\eeqa
where the corrections  $\Delta \vp , \Delta g _a, \Delta b _a$ contain some  contractions of $\nabla\vp, e^{\vp/2}V, e^{-\vp/2}W,\bH$  at order $\alpha'$. Note that the base space  metric, dilaton, $\bb$-field and the unite vector $n^a$ remain invariant at order $\alpha'$. Since the transformations must form the $Z_2$-group, the corrections  satisfy the following relations  \cite{Garousi:2019wgz}:
\beqa
-\Delta\vp (\psi)+\Delta\vp (\psi_0') &=&0\nn\\
\Delta b_a (\psi)+\Delta g_a (\psi'_0) &=&0\nn\\
\Delta g_a (\psi)+\Delta b_a (\psi'_0)&=&0\nn\\
\Delta \bH_{abc} (\psi)+\Delta \bH_{abc} (\psi'_0) &=&0\labell{Z22}
\eeqa
Then the  corrections  should  have the following terms:
 \beqa\Delta\vp&=&e_3\prt_a\vp\prt^a\vp+e_1\bH^2+e_2(e^\vp V^2+e^{-\vp}W^2)\nn\\
\Delta g_a &=&e_4e^{\vp/2}\bH_{abc} V^{bc}+e_5e^{-\vp/2}\prt^b\vp W_{ab}\nn\\
\Delta b_a &=&-e_4e^{-\vp/2}\bH_{abc} W^{bc}+e_5e^{\vp/2}\prt^b\vp V_{ab}\labell{DDD}
\eeqa
where $e_1,\cdots, e_5$ are some parameters that should be fixed by  the $Z_2$-symmetry of the effective actions. Note that we did not include the corrections which depend on the derivative of the base space dilaton and metric. The correction $\Delta \bH_{abc}$ is related to the corrections  $\Delta g_a$,  $\Delta b_a$ through the following relation  which is resulted from the Bianchi identity \reef{anB}:
\beqa
\Delta\bH_{abc}&=&\tilde H_{abc}-3e^{-\vp/2}W_{[ab}\Delta b_{c]}-3e^{\vp/2}\Delta g_{[a}V_{bc]}
\eeqa
where $\tilde H_{abc}$ is a $U(1)\times U(1)$ gauge invariant closed 3-form at order $\alpha'$ which is odd under parity.  It has the following terms:
\beqa
\tilde H_{abc}&=&e_6\prt_{[a}W_{b}{}^dV_{c]d}+e_7\prt_{[a}\bH_{bc]d}\nabla^d\vp\labell{tH}
\eeqa
where $e_6,e_7$ are two other parameters that the $Z_2$-symmetry of the effective action  should fix them.

 Using the reduction \reef{S0r}, then one can calculate $\Delta S_0$ from the expansion \reef{DS0} in terms of   above corrections, \ie
\beqa
\Delta S_0&=& -\frac{2 }{\kappa^2}\int d^{D-1}x e^{-2\bphi}\sqrt{-\bg} \,  \Big[ 
\frac{1}{4}\Big(  e^\vp V^2-e^{-\vp}W^2\Big)\Delta\vp\nn\\
 &&+\frac{1}{2}e^{-\vp/2}\prt_b\vp W^{ab}\Delta g_a -\frac{1}{2}e^{\vp/2}\prt_b\vp V^{ab}\Delta b_a -\frac{1}{6}\bH^{abc}\Delta\bH_{abc}\nn\\
 &&+\frac{1}{2}(\prt_a\vp+4\prt_a\bphi)\nabla^a(\Delta\vp)-\nabla_a\nabla^a(\Delta\vp)
 \nn\\
&&+e^{-\vp/2}W_{ab}\nabla^b(\Delta g^a)+e^{\vp/2}V_{ab}\nabla^b(\Delta b^a)\Big] \labell{delS}
 \eeqa
where no integration by part has been used. One can check that up to some total derivative terms the above expression become the same as the corresponding expression in \cite{Garousi:2019wgz}  for $\Delta\bphi=\Delta\bar{g}_{ab}=0$ (\ie eq.(23) in \cite{Garousi:2019wgz}), in which the  total derivative terms were ignored. Note  that $\Delta S_0$ is not even or odd under the Buscher rules. However, up to some total derivative terms it is odd under the Buscher rules \cite{Garousi:2019wgz}.

Using the reductions \reef{reduc}, it is straightforward to reduce the effective action $\!\!\bS_1$  with the Lagrangian density \reef{L1bulk} to find $S_1(\psi)$, and then calculate its transformation under the Buscher rules \reef{T2}, \ie $S_1(\psi_0')$. See \cite{Garousi:2019wgz}, for the details of this calculation. Note  that  $S_1(\psi)-S_1(\psi_0')$ is odd under the Buscher rules, however,  since  $\Delta S_0$ is not even or odd,  the vector $A_1^a(\psi)$  in \reef{S1b} is not even or odd under the Buscher rules either. Hence,
 the vector $A_1^a(\psi)$  should contain all  even-parity contractions of $\prt\vp$, $\prt\bphi$,  $e^{\vp/2}V, e^{-\vp/2}W$, $\bH$, $\bar{R}$ and their  derivatives at order $\alpha'$ with  unknown coefficients. Using the package ''xAct'', one finds it has 48 terms, \ie
\beqa
A_1^a&=&j_{1} \bH_{bcd} V^{bc} W^{ad} + j_{2} \bH^{a}{}_{cd} V^{bc}
W_{b}{}^{d} + j_{3} \bH_{bcd} V^{ab} W^{cd} + j_{4} \bH^{bcd}
\nabla^{a}\bH_{bcd} + j_{7} \nabla^{a}R\nn\\&& + j_{9}
\bH_{bcd} \bH^{bcd} \nabla^{a}\bphi + j_{11} R
\nabla^{a}\bphi + e^{\vp} j_{13} V_{bc} V^{bc}
\nabla^{a}\bphi + e^{-\vp} j_{15} W_{bc} W^{bc} \nabla^{a}\bphi \nn\\&&+
e^{\vp} j_{17} V^{bc} \nabla^{a}V_{bc} + e^{-\vp} j_{20} W^{bc}
\nabla^{a}W_{bc} + j_{23} \bH_{bcd} \bH^{bcd} \nabla^{a}\vp +
j_{25} R \nabla^{a}\vp \nn\\&&+ e^{\vp} j_{27}
V_{bc} V^{bc} \nabla^{a}\vp + e^{-\vp} j_{29} W_{bc} W^{bc}
\nabla^{a}\vp + j_{31}
\nabla^{a}\nabla_{b}\nabla^{b}\bphi + j_{33}
\nabla^{a}\nabla_{b}\nabla^{b}\vp\nn\\&& + j_{8}
\nabla_{b}R^{ab} + j_{41} \nabla^{a}\bphi
\nabla_{b}\nabla^{b}\bphi + j_{43} \nabla^{a}\vp
\nabla_{b}\nabla^{b}\bphi + j_{45} \nabla^{a}\bphi
\nabla_{b}\nabla^{b}\vp + j_{47} \nabla^{a}\vp
\nabla_{b}\nabla^{b}\vp\nn\\&& + j_{32}
\nabla_{b}\nabla^{b}\nabla^{a}\bphi + j_{34}
\nabla_{b}\nabla^{b}\nabla^{a}\vp + j_{10} \bH^{acd}
\bH_{bcd} \nabla^{b}\bphi + j_{12} R^{a}{}_{b}
\nabla^{b}\bphi + e^{\vp} j_{14} V^{ac} V_{bc}
\nabla^{b}\bphi \nn\\&&+ e^{-\vp} j_{16} W^{ac} W_{bc} \nabla^{b}\bphi +
j_{35} \nabla^{a}\bphi \nabla_{b}\bphi \nabla^{b}\bphi +
j_{36} \nabla^{a}\vp \nabla_{b}\bphi \nabla^{b}\bphi +
j_{37} \nabla^{a}\bphi \nabla_{b}\vp \nabla^{b}\bphi\nn\\&& +
j_{38} \nabla^{a}\vp \nabla_{b}\vp \nabla^{b}\bphi +
j_{42} \nabla_{b}\nabla^{a}\bphi \nabla^{b}\bphi + j_{46}
\nabla_{b}\nabla^{a}\vp \nabla^{b}\bphi + j_{24} \bH^{acd}
\bH_{bcd} \nabla^{b}\vp + j_{26} R^{a}{}_{b}
\nabla^{b}\vp \nn\\&&+ e^{\vp} j_{28} V^{ac} V_{bc}
\nabla^{b}\vp + e^{-\vp} j_{30} W^{ac} W_{bc} \nabla^{b}\vp +
j_{39} \nabla^{a}\bphi \nabla_{b}\vp \nabla^{b}\vp +
j_{40} \nabla^{a}\vp \nabla_{b}\vp \nabla^{b}\vp\nn\\&& +
j_{44} \nabla_{b}\nabla^{a}\bphi \nabla^{b}\vp + j_{48}
\nabla_{b}\nabla^{a}\vp \nabla^{b}\vp + e^{\vp}
j_{18} V^{bc} \nabla_{c}V^{a}{}_{b} + e^{\vp} j_{19}
V^{ab} \nabla_{c}V_{b}{}^{c}\nn\\&& + e^{-\vp} j_{21} W^{bc}
\nabla_{c}W^{a}{}_{b} + e^{-\vp} j_{22} W^{ab} \nabla_{c}W_{b}{}^{c} +
j_{5} \bH^{bcd} \nabla_{d}\bH^{a}{}_{bc} + j_{6} \bH^{abc}
\nabla_{d}\bH_{bc}{}^{d}
\eeqa
where $j_1,\cdots, j_{48}$ are 48 parameters. Note that some of the above terms are related by the Bianchi identities. We will remove them after imposing the constraint \reef{S1b}.

To find the parameters which satisfy the bulk constraint \reef{S1b}, one  should write the covariant derivatives and the curvatures  in \reef{S1b} in terms of partial derivative of the base space metric, and write the partial derivatives of the base space field strength $\bH, W,V$ in terms of the  potentials $\bb_{ab},g_a,b_a$ and then goes to  the local frame in which the first derivative of the metric is zero \cite{Garousi:2019cdn}. The coefficients of the resulting independent terms in the local frame then should be zero. This gives  some  linear equations involving all the parameters. We find the following solution in terms of the parameter $a_1,a_{12}$:
\beqa
&& a_{ 10}=-16 a_{ 1},a_{ 11}=0,a_{ 13}=384 a_{ 1}-12 a_{ 12},a_{ 14}=768 a_{ 1}-24 a_{ 12},a_{ 15}=96 a_{ 1}-6 a_{ 12},\nn\\&& a_{ 16}=-768 a_{ 1}+24 a_{ 12},a_{ 17}=-768 a_{ 1}+24 a_{ 12},a_{ 18}=48 a_{ 1},a_{ 19}=0,a_{ 2}=-9 a_{ 1}+(3 a_{ 12})/8, \nn\\&& a_{ 20}=-24 a_{ 1}+(3 a_{ 12})/2,a_{ 3}=a_{ 1}/3-a_{ 12}/96,a_{ 4}=72 a_{ 1}-3 a_{ 12},a_{ 5}=-192 a_{ 1}+6 a_{ 12}, \nn\\&& a_{ 6}=-8 a_{ 1}+a_{ 12}/4,a_{ 7}=48 a_{ 1}-(3 a_{ 12})/2,a_{ 8}=24 a_{ 1},a_{ 9}=-12 a_{ 1},e_{ 6}=-288 a_{ 1},e_{ 7}=0, \nn\\&& e_{ 1}=0,e_{ 2}=24 a_{ 1},e_{ 3}=48 a_{ 1},e_{ 4}=24 a_{ 1},e_{ 5}=48 a_{ 1},j_{ 1}=-24 a_{ 1},j_{ 10}=0,j_{ 11}=0,\nn\\&&  j_{ 12}=0,j_{ 13}=-48 a_{ 1},j_{ 14}=96 a_{ 1},j_{ 15}=48 a_{ 1},j_{ 16}=-96 a_{ 1},j_{ 17}=-48 a_{ 1},j_{ 18}=48 a_{ 1}, \nn\\&& j_{ 19}=-48 a_{ 1},j_{ 2}=0,j_{ 20}=-48 a_{ 1},j_{ 21}=-48 a_{ 1},j_{ 22}=48 a_{ 1},j_{ 23}=-8 a_{ 1}+a_{ 12}/2, \nn\\&& j_{ 24}=48 a_{ 1}-3 a_{ 12},j_{ 25}=192 a_{ 1}-6 a_{ 12},j_{ 26}=-384 a_{ 1}+12 a_{ 12},j_{ 27}=-72 a_{ 1}+(3 a_{ 12})/2, \nn\\&& j_{ 28}=192 a_{ 1}-6 a_{ 12},j_{ 29}=-24 a_{ 1}+(3 a_{ 12})/2,j_{ 3}=24 a_{ 1},j_{ 30}=192 a_{ 1}-6 a_{ 12},j_{ 31}=0, \nn\\&& j_{ 32}=0,j_{ 33}=0,j_{ 34}=0,j_{ 35}=0,j_{ 36}=384 a_{ 1}-12 a_{ 12},j_{ 37}=-768 a_{ 1}+24 a_{ 12},j_{ 38}=0, \nn\\&& j_{ 39}=0,j_{ 4}=0,j_{ 40}=24 a_{ 1}-(3 a_{ 12})/4,j_{ 41}=0,j_{ 42}=0,j_{ 43}=0,j_{ 44}=0,j_{ 45}=0,j_{ 46}=0,\nn\\&&  j_{ 47}=0,j_{ 48}=-96 a_{ 1},j_{ 5}=0,j_{ 6}=0,j_{ 8}=-2 j_{ 7},j_{ 9}=0\labell{sol4}
\eeqa
When replacing them into \reef{DDD} and \reef{tH}, one finds the following corrections to the Buscher rules:
 \beqa
  \Delta\vp&=&24 a_1\Big(2\prt_a\vp\prt^a\vp+e^\vp V^2+e^{-\vp}W^2\Big) \nn\\
  \Delta g_{a}&=&24 a_1\Big(2e^{-\vp/2}\prt^b\vp W_{ab}+e^{\vp/2}\bH_{abc} V^{bc} \Big)\nn\\
   \Delta b_{a}&=&24 a_1\Big(2e^{\vp/2}\prt^b\vp V_{ab}-e^{-\vp/2}\bH_{abc} W^{bc} \Big)\nn\\
   \Delta\bH_{abc}&=&-288a_1 \prt_{[a}(W_b{}^d V_{c]d})-3e^{\vp/2} V_{[ab}\Delta g_{c]}-3e^{-\vp/2} W_{[ab}\Delta b_{c]}\labell{dbH2}
  \eeqa
These transformations are  those  have been found in \cite{Kaloper:1997ux}\footnote{The sign of the first term in the last line of \reef{dbH2}, however, is different than the one appears in \cite{Kaloper:1997ux} which is a typo \cite{Borsato:2020bqo}.} for $24a_1=-\lambda_0$.

 Since the above corrections are independent of the parameter $a_{12}$, the solution \reef{sol4} produces  two mutliplets. One with the overall coefficient $a_1$ which is invariant under  the Buscher rules plus the above  higher derivative corrections, and the other one with the overall coefficient $a_{12}$ which is invariant under the Buscher rules.  The T-duality then can not fix a relation between these two parameters. We find the relation between them by using the fact that the sphere-level S-matrix elements of massless vertex operators in the string theory have simple poles reflecting the standard propagators for the massless and massive fields in the amplitude. At the low energy, one expands the massive propagators to find an amplitude in terms of only massless fields. The massless poles at the low energy are still simple poles. They should be reproduced by  effective actions which have standard massless propagators. The last term in the effective action \reef{L1bulk} changes the standard propagators of the $B$-field. However, for the following relation between $a_1$ and $a_{12}$:
\beqa
a_{12}&=&16a_1\labell{a112}
\eeqa
the coefficient of this term become zero after imposing the T-duality constraint \reef{sol4}, \ie $a_{20}=0$. Interestingly, imposing this relation, one finds the curvature terms in the action \reef{L1bulk} also becomes proportional to the Gauss-Bonnet gravity which does not change the standard propagator of the metric.

Imposing the relation \reef{a112} and the $Z_2$-constraint \reef{sol4}, one finds the following bulk action:
\beqa
\mathcal{L}_1&= &24a_1\Big[\frac{1}{24} H_{\alpha }{}^{\delta \epsilon } H^{\alpha \beta
\gamma } H_{\beta \delta }{}^{\varepsilon } H_{\gamma \epsilon
\varepsilon } -  \frac{1}{8} H_{\alpha \beta }{}^{\delta }
H^{\alpha \beta \gamma } H_{\gamma }{}^{\epsilon \varepsilon }
H_{\delta \epsilon \varepsilon } + \frac{1}{144} H_{\alpha
\beta \gamma } H^{\alpha \beta \gamma } H_{\delta \epsilon
\varepsilon } H^{\delta \epsilon \varepsilon }\nn\\&& + H_{\alpha }{}^{
\gamma \delta } H_{\beta \gamma \delta } R^{\alpha
\beta } - 4 R_{\alpha \beta } R^{\alpha \beta
} -  \frac{1}{6} H_{\alpha \beta \gamma } H^{\alpha \beta
\gamma } R + R^2 + R_{\alpha \beta
\gamma \delta } R^{\alpha \beta \gamma \delta } -
\frac{1}{2} H_{\alpha }{}^{\delta \epsilon } H^{\alpha \beta
\gamma } R_{\beta \gamma \delta \epsilon }\nn\\&& -
\frac{2}{3} H_{\beta \gamma \delta } H^{\beta \gamma \delta }
\nabla_{\alpha }\nabla^{\alpha }\Phi + \frac{2}{3} H_{\beta
\gamma \delta } H^{\beta \gamma \delta } \nabla_{\alpha }\Phi
\nabla^{\alpha }\Phi + 8 R \nabla_{\alpha }\Phi
\nabla^{\alpha }\Phi + 16 \nabla_{\alpha }\Phi \nabla^{\alpha
}\Phi \nabla_{\beta }\nabla^{\beta }\Phi \nn\\&&- 16
R_{\alpha \beta } \nabla^{\alpha }\Phi \nabla^{\beta
}\Phi - 16 \nabla_{\alpha }\Phi \nabla^{\alpha }\Phi \nabla_{
\beta }\Phi \nabla^{\beta }\Phi + 2 H_{\alpha }{}^{\gamma
\delta } H_{\beta \gamma \delta } \nabla^{\beta
}\nabla^{\alpha }\Phi \Big]\labell{fL1}
\eeqa
Since all background independent parameters in the gauge invariant action \reef{L1bulk} are fixed in the  particular geometry which has one circle, then the above action should be valid for any other geometry as well, \eg if one considers a geometry which has a tours $T^2$, then the reduction of the above action should have the symmetry $O(2,2)$.  The above action is exactly the off-shell Lagrangian that has been found in \cite{Meissner:1996sa} by imposing various field redefinitions on the on-shell action in the minimal scheme.

The relation \reef{a112} and the $Z_2$-constraint \reef{sol4} produces   the total derivative terms in \reef{S1b} that their corresponding  vector $A_1^a$ is the following:
\beqa
A_1^a&=&24a_1\Big[ \bH_{bcd} V^{ab} W^{cd} - \bH_{bcd} V^{bc} W^{ad}- 2 e^{\vp}
V_{bc} V^{bc} \nabla^{a}\bphi +2 e^{-\vp} W_{bc} W^{bc} \nabla^{a}\bphi - 2 e^{\vp} V^{bc} \nabla^{a}V_{bc}\nn\\&& -
2 {e^{-\vp}}W^{bc} \nabla^{a}W_{bc} + 4 R \nabla^{a}
\vp - 2 e^{\vp} V_{bc} V^{bc} \nabla^{a}\vp + 4 e^{\vp}
V^{ac} V_{bc} \nabla^{b}\bphi -  4 {e^{-\vp}}W^{ac} W_{bc} \nabla^{b}\bphi \nn\\&&+ 8 \nabla^{a}\vp \nabla_{b}\bphi \nabla^{b}\bphi - 16 \nabla^{a}\bphi \nabla_{b}\vp \nabla^{b}\bphi -
8 R^{a}{}_{b} \nabla^{b}\vp + 4 e^{\vp} V^{ac} V_{bc} \nabla^{b}\vp + 4 {e^{-\vp}}W^{ac} W_{bc} \nabla^{b}\vp
\nn\\&&+ \frac{1}{2} \nabla^{a}\vp \nabla_{b}\vp \nabla^{b}\vp - 4
\nabla_{b}\nabla^{a}\vp \nabla^{b}\vp + 2 e^{\vp} V^{bc} \nabla_{c}V^{a}{}_{b} - 2 e^{\vp} V^{ab} \nabla_{c}V_{b}{}^{c} -
2{e^{-\vp}} W^{bc} \nabla_{c}W^{a}{}_{b}\nn\\&& + 2 {e^{-\vp}}W^{ab} \nabla_{c}W_{b}{}^{c}\Big]\labell{A1}
\eeqa
Note that  the above vector is not odd or even under the Buscher rules.

The presence of the total derivative terms indicates that the bulk action alone is not invariant under the $Z_2$-transformations. Using the Stokes's theorem \reef{stok}, these anomalous terms are transferred to the boundary and  then  should be cancelled with the anomalous terms in the boundary action. The anomalous terms from the bulk action are then the following terms:
\beqa
&&\int_{\prt M}d^{D-2}\sigma\sqrt{|\tg|}n_{a} A_1^a e^{-2\bphi}=24 a_1\int_{\prt M}d^{D-2}\sigma\sqrt{|\tg|} e^{-2\bphi} n_{a}\Big[ \bH_{bcd} V^{ab} W^{cd} - \bH_{bcd} V^{bc} W^{ad}\nn\\&&- 2 e^{\vp}
V_{bc} V^{bc} \nabla^{a}\bphi +2 e^{-\vp} W_{bc} W^{bc} \nabla^{a}\bphi - 2 e^{\vp} V^{bc} \nabla^{a}V_{bc} -
2 {e^{-\vp}}W^{bc} \nabla^{a}W_{bc} + 4 R \nabla^{a}
\vp\nn\\&& - 2 e^{\vp} V_{bc} V^{bc} \nabla^{a}\vp + 4 e^{\vp}
V^{ac} V_{bc} \nabla^{b}\bphi -  4 {e^{-\vp}}W^{ac} W_{bc} \nabla^{b}\bphi + 8 \nabla^{a}\vp \nabla_{b}\bphi \nabla^{b}\bphi - 16 \nabla^{a}\bphi \nabla_{b}\vp \nabla^{b}\bphi\nn\\&& -
8 R^{a}{}_{b} \nabla^{b}\vp + 4 e^{\vp} V^{ac} V_{bc} \nabla^{b}\vp + 4 {e^{-\vp}}W^{ac} W_{bc} \nabla^{b}\vp
 + \frac{1}{2} \nabla^{a}\vp \nabla_{b}\vp \nabla^{b}\vp - 4
\nabla_{b}\nabla^{a}\vp \nabla^{b}\vp \nn\\&&+ 2 e^{\vp} V^{bc} \nabla_{c}V^{a}{}_{b} - 2 e^{\vp} V^{ab} \nabla_{c}V_{b}{}^{c} -
2{e^{-\vp}} W^{bc} \nabla_{c}W^{a}{}_{b} + 2 {e^{-\vp}}W^{ab} \nabla_{c}W_{b}{}^{c}\Big]\labell{A11}
\eeqa
In the next subsection we study the $Z_2$-transformations of the boundary action \reef{L1boundary}.

\subsection{Boundary constraint}

To impose the $Z_2$-constraint \reef{S11b} on the boundary terms, one   first assumes  that  the boundary is specified by the functions $x^\mu(\sigma^{\tilde{\mu}})=(x^a(\sigma^\ta),y)$ where the functions $x^a(\sigma^\ta)$ represent the boundary in the base space which are independent of the $y$-coordinate.  Then the reduction of the induce metric \reef{indg} becomes
\beqa
g_{\tilde{\mu}\tilde{\nu}}=\left(\matrix{\frac{\prt x^a}{\prt \sigma^\ta}\frac{\prt x^b}{\prt \sigma^\tb}(\bg_{a b}+e^{\varphi}g_{a }g_{b })& \frac{\prt x^a}{\prt \sigma^\ta}(e^{\varphi}g_{a })&\cr\frac{\prt x^b}{\prt \sigma^\tb}( e^{\varphi}g_{b })&e^{\varphi}&}\right)
\eeqa
Using this and the reduction of dilaton in \reef{reduc}, one finds the following reduction:
\beqa
e^{-2\Phi}\sqrt{| g|}&=& e^{-2\bphi}\sqrt{| \tg|}
\eeqa
where $\tg$ is the determinate of the induced metric \reef{gtatb}. For the scheme that we have considered  for the T-duality transformations, it is invariant under the $Z_2$-transformations to all orders of $\alpha'$.

Using the reduction \reef{Rn} for the normal vector and the reduction \reef{reduc} for metric, one finds the reduction of the extrinsic curvature to be
\beqa
G^{\mu\nu}K_{\mu\nu}&=& \bg^{ab}\bK_{ab}+\frac{1}{2}n^a\nabla_a\vp
\eeqa
where $\bK_{ab}$ is the extrinsic curvature of the boundary of the base space. So the reduction of the boundary action $\prt S_0$ in \reef{baction} is
\beqa
\prt S_0(\psi)&=&-\frac{4}{\kappa^2}\int d^{D-2}\sigma\,e^{-2\bphi}\sqrt{| \tg|}\Big[  \bg^{ab}\bK_{ab}+\frac{1}{2} n^a\nabla_a\vp\Big]\labell{redcebs}
\eeqa
which is invariant under $U(1)\times U(1)$ gauge transformations.

In using the boundary  constraint \reef{S11b}, one needs to calculate $\Delta \prt S_0$. Using the above  reduction, one can calculate $\Delta \prt S_0$ in terms of the  T-duality transformations \reef{T22}, \ie
\beqa
\Delta \prt S_0&=&-\frac{4}{\kappa^2}\int d^{D-2}\sigma\,e^{-2\bphi}\sqrt{| \tg|}\Big[ \frac{1}{2}n^a\nabla_a(\Delta\vp)\Big]
\eeqa
where $\Delta\vp$ is given in \reef{dbH2}. Note that in finding the above result we have used the fact that in the scheme that we have considered,  the unit vector $n^a$ and the base space metric and dilaton  are  invariant under the $Z_2$-transformations.

Using the reductions \reef{reduc}, it is straightforward to reduce the effective action $\prt\!\!\bS_1$  with the Lagrangian density \reef{L1boundary} to find $\prt S_1(\psi)$, and then calculate its transformation under the Buscher rules \reef{T2}, \ie $\prt S_1(\psi_0')$.   See \cite{Garousi:2019mca,Akou:2020mxx}, for  the reduction of different tensors in the boundary Lagrangian \reef{L1boundary}. Note that for simplicity of calculations in the boundary, we assume the metric of the base space is flat, \ie $\bg_{ab}=\eta_{ab}$. This calculation produces  the expression $\prt S_1(\psi)-\prt S_1(\psi_0')$ in \reef{S11b}. The anomalous couplings from the T-duality of the bulk action is also given in \reef{A11}.  Finally, one needs to add the total derivative terms $T_1(\psi)$  to the constraint  \reef{S11b}.  According to the Stokes's theorem, the total derivative terms in the boundary which have the following structure are zero:
\beqa
T_1(\psi)=-\frac{2}{\kappa^2}\int_{\prt M}d^{D-2}\sigma \sqrt{|\tg |} n_{a}\prt_{b}(e^{-2\bphi}F^{ab})
\labell{bstokes1}
\eeqa
where  $ F^{ab} $ is an arbitrary even-parity antisymmetric  tensor constructed  from $U(1)\times U(1)$ gauge invariant tensors   $n,\prt n,\prt\bphi,\prt\vp,e^{-\vp/2}W,e^{\vp/2}V,\bH$ at two derivative order. Using the package "xAct", one can construct this tensor with arbitrary  coefficients.

Having calculated all terms in \reef{S11b}, one should then impose the Bianchi identities \reef{anB} in the flat base space which can be done by writing the  derivatives of the field strengths $V,W,\bH$  in terms of potentials $g_a,b_a,\bb_{ab}$.  We consider in this section only  the timelike boundary for which $n^\mu n_\mu=1$ and $n^an_a=1$. To impose the identities corresponding to the unite vector $n^a$ in the base space, we also write it in terms of the function $f$ using \reef{nf}, \ie
\beqa
n^{a}=(\prt_b f\prt^b f)^{-1/2}\prt^a f\labell{nfa}
\eeqa
where we have used the fact that the function $f$ should be independent of the killing coordinate $y$, \ie $\prt_y f=0$.  The coefficients of the resulting independent terms then should be zero. They produce  some  linear equations involving  $a_1$ and the parameters in \reef{L1boundary} and in \reef{bstokes1}.

We find   the following non-zero couplings in the timelike boundary:
\beqa
\prt \cL_1&\!\!\!\!\!=\!\!\!\!\!&b_{1}^{} H_{\beta \gamma \delta } H^{\beta \gamma \delta }
K^{\alpha }{}_{\alpha } + 24 a_{1}^{} H_{\alpha }{}^{\gamma
\delta } H_{\beta \gamma \delta } K^{\alpha \beta } +
b_{11}^{} K_{\alpha }{}^{\gamma } K^{\alpha \beta } K_{\beta
\gamma } + (24 b_{1}^{} -  \frac{1}{2} b_{17}^{}) K^{\alpha
}{}_{\alpha } K_{\beta \gamma } K^{\beta \gamma }\nn\\&& + (-8
b_{1}^{} -  \frac{1}{6} b_{18}^{}) K^{\alpha }{}_{\alpha }
K^{\beta }{}_{\beta } K^{\gamma }{}_{\gamma } -  \frac{1}{2}
b_{19}^{} H_{\alpha }{}^{\delta \epsilon } H_{\beta \delta
\epsilon } K^{\gamma }{}_{\gamma } n^{\alpha } n^{\beta } +
12 (-16 a_{1}^{} + b_{12}^{}) K^{\alpha \beta }
R_{\alpha \beta } \nn\\&&+ (96 a_{1}^{} + 24 b_{1}^{} - 12
b_{12}^{} -  \frac{1}{2} b_{17}^{} + 2 b_{19}^{}) K^{\gamma }{}_{
\gamma } n^{\alpha } n^{\beta } R_{\alpha \beta } -
12 b_{1}^{} K^{\alpha }{}_{\alpha } R + b_{11}^{}
K^{\gamma \delta } n^{\alpha } n^{\beta } R_{\alpha
\gamma \beta \delta }\nn\\&& + b_{12}^{} H^{\beta \gamma \delta } n^{
\alpha } \nabla_{\alpha }H_{\beta \gamma \delta } - 2 b_{1}^{}
H_{\beta \gamma \delta } H^{\beta \gamma \delta } n^{\alpha }
\nabla_{\alpha }\Phi + b_{17}^{} K_{\beta \gamma } K^{\beta
\gamma } n^{\alpha } \nabla_{\alpha }\Phi + b_{18}^{}
K^{\beta }{}_{\beta } K^{\gamma }{}_{\gamma } n^{\alpha }
\nabla_{\alpha }\Phi \nn\\&&+ b_{19}^{} H_{\beta }{}^{\delta \epsilon
} H_{\gamma \delta \epsilon } n^{\alpha } n^{\beta }
n^{\gamma } \nabla_{\alpha }\Phi + (-192 a_{1}^{} + 24 b_{12}^{}
+ b_{17}^{} - 4 b_{19}^{}) n^{\alpha } n^{\beta } n^{\gamma }
R_{\beta \gamma } \nabla_{\alpha }\Phi \nn\\&&+ 24 (8
a_{1}^{} + b_{1}^{}) n^{\alpha } R \nabla_{\alpha
}\Phi - 48 b_{1}^{} K^{\beta }{}_{\beta } \nabla_{\alpha }\Phi
\nabla^{\alpha }\Phi - 2 (-192 a_{1}^{} - 48 b_{1}^{} + b_{18}^{})
K^{\gamma }{}_{\gamma } n^{\alpha } n^{\beta }
\nabla_{\alpha }\Phi \nabla_{\beta }\Phi\nn\\&& + 24 (-16 a_{1}^{} - 2
b_{1}^{} + b_{12}^{}) K^{\alpha \beta } \nabla_{\beta }\nabla_{
\alpha }\Phi + (192 a_{1}^{} - 24 b_{12}^{} -  b_{17}^{} + 4
b_{19}^{}) K^{\gamma }{}_{\gamma } n^{\alpha } n^{\beta }
\nabla_{\beta }\nabla_{\alpha }\Phi \nn\\&&+ 48 (-8 a_{1}^{} -
b_{1}^{}) n^{\alpha } R_{\alpha \beta } \nabla^{\beta
}\Phi - 384 a_{1}^{} K_{\alpha \beta } \nabla^{\alpha }\Phi
\nabla^{\beta }\Phi + 96 (4 a_{1}^{} + b_{1}^{}) n^{\alpha }
\nabla_{\alpha }\Phi \nabla_{\beta }\Phi \nabla^{\beta }\Phi
\nn\\&&+ 96 (-8 a_{1}^{} -  b_{1}^{}) n^{\alpha } \nabla_{\beta
}\nabla_{\alpha }\Phi \nabla^{\beta }\Phi + (48 a_{1}^{} -
\frac{1}{4} b_{11}^{} - 6 b_{12}^{}) H_{\alpha }{}^{\delta
\epsilon } n^{\alpha } n^{\beta } n^{\gamma } \nabla_{\gamma
}H_{\beta \delta \epsilon }\nn\\&&+ (384 a_{1}^{} + 96 b_{1}^{} + 48 b_{12}^{} + 2 b_{17}^{} -
8 b_{19}^{}) n^{\alpha } n^{\beta } n^{\gamma }
\nabla_{\alpha }\Phi \nabla_{\gamma }\nabla_{\beta }\Phi +
b_{38}^{} n^{\alpha } n^{\beta } n^{\gamma } n^{\delta }
\nabla_{\delta }\nabla_{\gamma }K_{\alpha \beta } \nn\\&&+ \frac{4}{3} (-576 a_{1}^{} - 96
b_{1}^{} + b_{18}^{}) n^{\alpha } n^{\beta } n^{\gamma }
\nabla_{\alpha }\Phi \nabla_{\beta }\Phi \nabla_{\gamma
}\Phi \labell{L12}
\eeqa
which has the bulk parameter $a_1$ and 7 boundary parameters $b_1,b_{11},b_{12},b_{17},b_{18},b_{19},b_{38}$.
The corresponding antisymmetric tensor  $ F^{ab} $ in the Stokes's theorem is
\beqa
 F^{ab}&\!\!\!\!\!=\!\!\!\!\!&6 (8 a_{1}^{} - b_{12}^{}) e^{\vp} n^{b} n^{c} V^{ad} V_{cd} -
6 (8 a_{1}^{} - b_{12}^{}) e^{\vp} n^{a} n^{c} V^{bd} V_{cd} -
6 (8 a_{1}^{} - b_{12}^{}) {e^{-\vp}}n^{b} n^{c} W^{ad}
W_{cd}\nn\\&& -  6 (-8 a_{1}^{} + b_{12}^{}) {e^{-\vp}}n^{a}
n^{c} W^{bd} W_{cd} + 24 b_{1}^{} n^{b}\prt^{a}\vp\prt_{c}n^{c} - 24 b_{1}^{} n^{a}\prt^{b}\vp\prt_{c}n^{c}\nn\\&& - 48 (8 a_{1}^{}+ b_{1}^{}) n^{b} n^{c} \prt^{a}\vp\prt_{c}\bphi + 48 (8 a_{1}^{} + b_{1}^{}) n^{a}
n^{c}\prt^{b}\vp\prt_{c}\bphi - 24 b_{1}^{} n^{b}\prt^{a}n^{c}\prt_{c}\vp + 24 b_{1}^{} n^{a}\prt^{b}n^{c} \prt_{c}\vp\nn
\eeqa
which has the bulk parameter $a_1$ and the boundary parameters $b_1,b_{12}$. The background independent parameters $b_1$, $b_{11}$, $b_{12}$, $b_{17}$, $b_{18}$, $b_{19}$, $b_{38}$  may be fixed further by considering the geometry which has the tours $T^2$ in which we are not interested in this paper. In this paper, however, we further constrain the parameters by using  the least action principle, and by  considering the geometry which has tours $T^d$. 
 Before imposing  these conditions, let us compare the gravity couplings that the $Z_2$-symmetry produces, with the  couplings in the Euler character $I_2$.

\subsubsection{Comparing with the  Euler character $I_2$}

In this subsection we compare the  couplings in the classical effective action that are consistent with the T-duality, with the  couplings in the Chern-Simons form $Q_2$ in the Euler character $I_2$.    When the dilaton and $B$-field are zero, the couplings that are found by the T-duality constraint are the following:
\beqa
\bS_1+\prt\!\!\bS_1&=&-\frac{48a_1}{\kappa^2}\int d^Dx\sqrt{-G}(R^{\mu\nu\alpha\beta}R_{\mu\nu\alpha\beta}-4R^{\mu\nu}R_{\mu\nu}+R^2)\nn\\&&
-\frac{2}{\kappa^2}\int d^{D-1}x\sqrt{|g|}\Bigg[
b_{11}^{} K_{\alpha }{}^{\gamma } K^{\alpha \beta } K_{\beta
\gamma } + (24 b_{1}^{} -  \frac{1}{2} b_{17}^{}) K^{\alpha
}{}_{\alpha } K_{\beta \gamma } K^{\beta \gamma }\nn\\&& + (-8
b_{1}^{} -  \frac{1}{6} b_{18}^{}) K^{\alpha }{}_{\alpha }
K^{\beta }{}_{\beta } K^{\gamma }{}_{\gamma } +
12 (-16 a_{1}^{} + b_{12}^{}) K^{\alpha \beta }
R_{\alpha \beta } \nn\\&&+ (96 a_{1}^{} + 24 b_{1}^{} - 12
b_{12}^{} -  \frac{1}{2} b_{17}^{} + 2 b_{19}^{}) K^{\gamma }{}_{
\gamma } n^{\alpha } n^{\beta } R_{\alpha \beta } -
12 b_{1}^{} K^{\alpha }{}_{\alpha } R\nn\\&& + b_{11}^{}
K^{\gamma \delta } n^{\alpha } n^{\beta } R_{\alpha
\gamma \beta \delta } +
b_{38}^{} n^{\alpha } n^{\beta } n^{\gamma } n^{\delta }
\nabla_{\delta }\nabla_{\gamma }K_{\alpha \beta } \Bigg]\labell{b11}
\eeqa
where the couplings are all independent.  One may use various identities to rewrite the boundary coupling in different form, however, the number of boundary couplings remain the same. For example, one may use the following identity:
\beqa
n^{\alpha } n^{\beta } n^{\gamma } n^{\delta }
\nabla_{\delta }\nabla_{\gamma }K_{\alpha \beta }&=&-2 K_{\alpha }{}^{\gamma } K^{\alpha \beta } K_{\beta
\gamma }+n^{\alpha } n^{\beta }
\nabla_{\gamma }\nabla^{\gamma }K_{\alpha \beta }\labell{iden}
\eeqa
which can be verified by writing both sides in terms of function $f$, to write the term with the second derivative of extrinsic curvature in \reef{b11} in terms of the Laplacian of the extrinsic curvature. This  changes the coefficient of $K_{\alpha }{}^{\gamma } K^{\alpha \beta } K_{\beta
\gamma }$, however, the number of the boundary couplings remain the same.

On the other hand, it has been shown in \cite{Myers:1987yn} that if one extends the couplings in the 4-dimensional  Euler density $I_2$ to arbitrary dimension, the couplings satisfy  the least action principle with  the boundary condition that only metric is  arbitrary on the boundary. This extension is
\beqa
I_2&=&\int d^Dx\sqrt{-G} R^2_{\rm GB}+\int d^{D-1}x\sqrt{-g}Q_2\labell{Euler}
\eeqa
where $R^2_{\rm GB}$ is the Gauss-Bonnet gravity and $Q_2$ is the Chern-Simons form for the timelike boundary \cite{Myers:1987yn}
\beqa
Q_2&=&4\Bigg[K^\mu{}_{\mu}\tR-2K^{\mu\nu}\tR_{\mu\nu}+\frac{1}{3}(3K^\alpha{}_{\alpha}K_{\mu\nu}K^{\mu\nu}-K^\mu{}_{\mu}K^\nu{}_{\nu}K^\alpha{}_{\alpha}-2K_{\mu}{}^{\nu}K_{\nu\alpha}K^{\alpha\mu})\Bigg]
\eeqa
where $\tR_{\mu\nu}$ and $\tR$ are curvatures that are constructed from the induced metric \reef{indg}. Using the  following Gauss-Codazzi  relations for the timelike boundary:
 \beqa
\tR_{\alpha\beta}&=&P_{\alpha\mu}P_{\beta\nu}R^{\mu\nu}-n^\mu n^\nu R_{\alpha\mu\beta\nu}-K_{\alpha\mu}K_{\beta}{}^{\mu}+K_{\alpha\beta}K_\mu {}^{\mu}\nn\\
\tR&=&R-2n^\mu n^\nu R_{\mu\nu}-K_{\mu\nu}K^{\mu\nu}+K_\mu {}^\mu K_\nu {}^\nu
\eeqa
and the identity $n^\mu K_{\mu\nu}=0$, one can rewrite $Q_2$ in terms of the spacetime curvatures, \ie
\beqa
Q_2&=&4\Bigg[K^\mu{}_{\mu}R-2K^{\mu\nu}R_{\mu\nu}-2K_\alpha{}^\alpha n^\mu n^\nu R_{\mu\nu}+2K^{\mu\nu}n^\alpha n^\beta R_{\alpha\mu\beta\nu}\labell{I4}\\&&\qquad-\frac{1}{3}(6K^\alpha{}_{\alpha}K_{\mu\nu}K^{\mu\nu}-2K^\mu{}_{\mu}K^\nu{}_{\nu}K^\alpha{}_{\alpha}-4K_{\mu}{}^{\nu}K_{\nu\alpha}K^{\alpha\mu})\Bigg]\nn
\eeqa
The bulk couplings in $I_2$ are the same as the bulk couplings in \reef{b11}, however, the number of boundary couplings in \reef{b11} is more that the number of boundary couplings in $I_2$. If one sets $b_{38}$ to zero in \reef{b11}, then the number of couplings  and the structure of couplings become the same as those in $I_2$, however, for no values of the boundary parameters $b_1, b_{11}, b_{12}, b_{17}, b_{18}, b_{19}$ the two sets of the couplings become identical, \eg  the ratio of  the last terms in the first and second lines above is $3/2$ whereas this ratio in \reef{b11} is one. Hence, the boundary couplings that the T-duality  dictates can not be ${\it exactly}$ the same as  the boundary couplings in  $I_2$ for any values of the boundary parameters.  This may indicate that the assumption that only metric is arbitrary on the boundary in the 4-derivative couplings is not consistent with the classical effective action of string theory. However, such assumption may be valid for the higher loop effective actions for which there is no T-duality symmetry. 

\section{Constraint from the least action principle}

We have seen that when the background geometry has a circle, requiring the bulk and boundary actions at order $\alpha'$ to have the $Z_2$-symmetry, one finds the bulk action \reef{fL1} and the boundary action \reef{L12}. Using the background independent assumption, then these actions should be the effective actions  at order $\alpha'$ for any arbitrary background which has timelike boundary, up to field redefinitions.  To find the  bulk equations of motion one needs to extremize these actions,  \ie $\delta(\!\!\bS_1+\prt\!\!\bS_1)=0$, by using some assumption for the massless fields on the boundary. At the two derivative order, the assumption that the massless fields should be arbitrary on the boundary is good enough to produce the equations of motion from extremizing the leading order actions, \ie $\delta(\!\!\bS_0+\prt\!\!\bS_0)=0$. At the four derivative order, however,  the assumption that only the massless fields are arbitrary  requires the boundary couplings which are not consistent with the T-duality symmetry. If one insists on  that assumption, then the 4-derivative action would not be the classical effective action of the string theory in which the T-duality is a symmetry. Hence, it seems to study the classical equations of motion in the string theory at order $\alpha'$, one should assume that not only the massless fields but also their first derivatives should be arbitrary on the boundary.

Since the bulk action \reef{fL1} has at most two-derivative terms, \eg $R$ or $\nabla\nabla\Phi$, when one extremizes the bulk action, one would find the variation $\delta \Psi$ as well as the   variations $\nabla(\delta \Psi)$ and $\nabla\nabla(\delta\Psi)$ where $\Psi$ represents the massless fields. After using the Stokes's theorem, the   variations $\nabla(\delta \Psi)$ and $\nabla\nabla(\delta\Psi)$  produce the variations $\delta \Psi$ and $\nabla(\delta \Psi)$ on the boundary. The assumption that  $\Psi$ and $\nabla\Psi$ are arbitrary on the boundary, means their variations are zero on the boundary, \ie $\delta\Psi=\nabla(\delta \Psi)=0$. Hence, the bulk action satisfies $\delta\!\!\bS_1=(\cdots)\delta \Psi=0$ with no constraint on the bulk couplings.  The boundary action \reef{L12}, however, has two- and three-derivative terms, \eg $R$ or $\nabla\nabla K$. When one extremizes the boundary action, one would find variations  $\nabla\nabla(\delta \Psi)$ and $\nabla\nabla\nabla(\delta \Psi)$ on the boundary which are not zero in general. The parameters in the boundary action should be such that the coefficients of these variations become zero up to some total derivative terms on the boundary which are zero according to the Stokes's theorem.

The variation of the boundary action \reef{L12} against  the metric variation produces the following non-zero terms in the local frame:
\beqa
&&24 (8 a_{1}^{} + b_{1}^{}) \prt^{\alpha }\Phi f1^{\beta }
P^{\gamma \delta } \partial_{\alpha }\partial_{\beta
}\delta G_{\gamma \delta } + (96 a_{1}^{} -  \frac{1}{2} b_{11}^{} - 6
b_{12}^{}) f1^{\alpha } f1^{\beta } f2^{\gamma \delta }
P_{\gamma }{}^{\epsilon } P_{\delta }{}^{\varepsilon }
\partial_{\alpha }\partial_{\beta }\delta G_{\epsilon \varepsilon }\nn\\&& +
(-48 a_{1}^{} + 6 b_{12}^{} + \frac{1}{4} b_{17}^{} -  b_{19}^{})
f1^{\alpha } f1^{\beta } f2^{\gamma \delta } P_{\gamma
\delta } P^{\epsilon \varepsilon } \partial_{\alpha }\partial_{
\beta }\delta G_{\epsilon \varepsilon }  
\nn\\&&+ (-96 a_{1}^{} - 24 b_{1}^{} - 12 b_{12}^{} -
 \frac{1}{2} b_{17}^{} + 2 b_{19}^{}) \prt^{\alpha }\Phi
f1_{\alpha } f1^{\beta } f1^{\gamma } P^{\delta \epsilon }
\partial_{\beta }\partial_{\gamma }\delta G_{\delta \epsilon }\labell{varg}
\eeqa
where $f1^\alpha=\prt^\alpha f, f2^{\alpha\beta}=\prt^\alpha\prt^\beta f$. We have used the assumption that the variation of metric and its first derivative, and their tangent derivatives   are zero, \ie $\delta G_{\alpha\beta}=\prt_\mu\delta G_{\alpha\beta}=0$ and $P^{\mu\nu}\prt_\mu\prt_\gamma\delta G_{\alpha\beta}=0$. One is free to add arbitrary total derivative terms, \ie \reef{totb} in which the antisymmetric tensor $\cF_1^{\alpha\beta}$ contains the variation of metric. Using the Stokes's theorem, the total derivative terms on the boundary become zero. Then, up to some total derivative terms, the resulting equations are zero for the following relation:
\beqa
b_{19}= -48 a_1 +6 b_{12} +
  b_{17}/4\labell{re1}
\eeqa
Note that if one requires that the first derivative of the variation of metric to be non-zero on the boundary, \ie $\prt_\mu\delta G_{\alpha\beta}\neq 0$, then one would find many other terms in \reef{varg} that become zero for the incorrect value of  $a_1=0$.

 Inserting the relation \reef{re1} into \reef{L12}, one finds the variation of \reef{L12} against  the dilaton becomes zero up to some total derivative terms,  and the $B$-field variation produces the following relation:
 \beqa
 b_{12}=  16 a_1 - b_{11}/12,
\labell{re12}
\eeqa
and some total derivative terms. 
Therefore, for the relations \reef{re1} and \reef{re12}  the bulk and boundary actions satisfy the stationary condition  $\delta(\!\!\bS_1+\prt\!\!\bS_1)=(\cdots)\delta \Psi=0$ when the variation of massless fields and their first derivatives on the boundary are zero\footnote{If one does not use the total derivative terms \reef{totb} in the metric variation of the boundary couplings \reef{varg}, then one would find the constraint \reef{re1}, \reef{re12}, and another constraint $b_1=-8a_1$. }. Since the variation of fields in the bulk are non-zero, this gives the appropriate equations of motion in which we are not interested.

Inserting   the relations \reef{re1} and \reef{re12} into the boundary action \reef{L12}, one  finds the boundary action to be
\beqa
\prt \cL_1&\!\!\!\!=\!\!\!\!& b_{11}\Bigg[K_{\alpha }{}^{\gamma } K^{\alpha \beta }
K_{\beta \gamma } + \frac{1}{4} H_{\alpha }{}^{\delta
\epsilon } H_{\beta \delta \epsilon } K^{\gamma }{}_{\gamma }
n^{\alpha } n^{\beta } -  K^{\alpha \beta }
R_{\alpha \beta } + K^{\gamma \delta } n^{\alpha }
n^{\beta } R_{\alpha \gamma \beta \delta }\nn\\&& -
\frac{1}{12} H^{\beta \gamma \delta } n^{\alpha }
\nabla_{\alpha }H_{\beta \gamma \delta } -  \frac{1}{2}
H_{\beta }{}^{\delta \epsilon } H_{\gamma \delta \epsilon }
n^{\alpha } n^{\beta } n^{\gamma } \nabla_{\alpha }\Phi - 2
K^{\alpha \beta } \nabla_{\beta }\nabla_{\alpha }\Phi +
\frac{1}{4} H_{\alpha }{}^{\delta \epsilon } n^{\alpha }
n^{\beta } n^{\gamma } \nabla_{\gamma }H_{\beta \delta
\epsilon }\Bigg]\nn\\&&+ b_{1}^{} \Bigg[H_{\beta \gamma \delta } H^{\beta \gamma \delta }
K^{\alpha }{}_{\alpha } + 24 K^{\alpha }{}_{\alpha } K_{\beta
\gamma } K^{\beta \gamma } - 8 K^{\alpha }{}_{\alpha }
K^{\beta }{}_{\beta } K^{\gamma }{}_{\gamma } + 24 K^{\gamma
}{}_{\gamma } n^{\alpha } n^{\beta } R_{\alpha \beta
} - 12 K^{\alpha }{}_{\alpha } R\nn\\&& - 2 H_{\beta \gamma
\delta } H^{\beta \gamma \delta } n^{\alpha } \nabla_{\alpha
}\Phi + 24 n^{\alpha } R \nabla_{\alpha }\Phi - 48
K^{\beta }{}_{\beta } \nabla_{\alpha }\Phi \nabla^{\alpha
}\Phi + 96 K^{\gamma }{}_{\gamma } n^{\alpha } n^{\beta }
\nabla_{\alpha }\Phi \nabla_{\beta }\Phi \nn\\&&- 48 K^{\alpha
\beta } \nabla_{\beta }\nabla_{\alpha }\Phi - 48 n^{\alpha }
R_{\alpha \beta } \nabla^{\beta }\Phi + 96 n^{\alpha
} \nabla_{\alpha }\Phi \nabla_{\beta }\Phi \nabla^{\beta
}\Phi - 96 n^{\alpha } \nabla_{\beta }\nabla_{\alpha }\Phi
\nabla^{\beta }\Phi \nn\\&&- 128 n^{\alpha } n^{\beta } n^{\gamma }
\nabla_{\alpha }\Phi \nabla_{\beta }\Phi \nabla_{\gamma
}\Phi + 96 n^{\alpha } n^{\beta } n^{\gamma } \nabla_{\alpha
}\Phi \nabla_{\gamma }\nabla_{\beta }\Phi\Bigg]\nn\\&& + a_{1}^{} \Bigg[24
H_{\alpha }{}^{\gamma \delta } H_{\beta \gamma \delta }
K^{\alpha \beta } - 24 H_{\alpha }{}^{\delta \epsilon }
H_{\beta \delta \epsilon } K^{\gamma }{}_{\gamma } n^{\alpha
} n^{\beta } + 16 H^{\beta \gamma \delta } n^{\alpha }
\nabla_{\alpha }H_{\beta \gamma \delta }\nn\\&& + 48 H_{\beta
}{}^{\delta \epsilon } H_{\gamma \delta \epsilon } n^{\alpha }
n^{\beta } n^{\gamma } \nabla_{\alpha }\Phi + 192 n^{\alpha
} R \nabla_{\alpha }\Phi + 384 K^{\gamma }{}_{\gamma }
n^{\alpha } n^{\beta } \nabla_{\alpha }\Phi \nabla_{\beta
}\Phi - 384 n^{\alpha } R_{\alpha \beta }
\nabla^{\beta }\Phi\nn\\&& - 384 K_{\alpha \beta } \nabla^{\alpha
}\Phi \nabla^{\beta }\Phi + 384 n^{\alpha } \nabla_{\alpha
}\Phi \nabla_{\beta }\Phi \nabla^{\beta }\Phi - 768
n^{\alpha } \nabla_{\beta }\nabla_{\alpha }\Phi
\nabla^{\beta }\Phi \nn\\&&- 48 H_{\alpha }{}^{\delta \epsilon }
n^{\alpha } n^{\beta } n^{\gamma } \nabla_{\gamma }H_{\beta
\delta \epsilon } - 768 n^{\alpha } n^{\beta } n^{\gamma }
\nabla_{\alpha }\Phi \nabla_{\beta }\Phi \nabla_{\gamma
}\Phi + 768 n^{\alpha } n^{\beta } n^{\gamma }
\nabla_{\alpha }\Phi \nabla_{\gamma }\nabla_{\beta }\Phi\Bigg]\nn\\&&
-\frac{b_{17}}{2} \Bigg[ K^{\alpha }{}_{\alpha } K_{\beta
\gamma } K^{\beta \gamma } +  \frac{1}{4} H_{\alpha
}{}^{\delta \epsilon } H_{\beta \delta \epsilon } K^{\gamma
}{}_{\gamma } n^{\alpha } n^{\beta } - K_{\beta \gamma } K^{
\beta \gamma } n^{\alpha } \nabla_{\alpha }\Phi -
\frac{1}{2} H_{\beta }{}^{\delta \epsilon } H_{\gamma \delta
\epsilon } n^{\alpha } n^{\beta } n^{\gamma } \nabla_{\alpha
}\Phi\Bigg]\nn\\&&
-\frac{b_{18}}{6}\Big[K^\alpha{}_\alpha-2n^\alpha\nabla_\alpha\Phi\Big]^3 + b_{38}^{} n^{\alpha } n^{\beta } n^{\gamma }
n^{\delta } \nabla_{\delta }\nabla_{\gamma }K_{\alpha \beta
}\labell{fbaction}
\eeqa
The boundary multiplets with the background independent parameters $b_1,b_{11}, b_{17}, b_{18}$ and $b_{38}$ are each invariant under the $Z_2$-symmetry and satisfy the least action.  The boundary multiplet with parameter $a_1$ also satisfies the least action principle, however, its combination with the bulk multiplet \reef{fL1} satisfies the $Z_2$-symmetry. In the next section we study another constraint on  the background independent parameters by considering the geometry which has the tours $T^d$. 

\section{Constraint from zero cosmological boundary action}

In this section we show that the cosmological/one-dimensional reduction of the boundary term at the leading order  produces zero boundary action. We then extend this to the cosmological/one-dimensional  boundary action at order $\alpha'$ to further constrain the boundary parameters in the timelike boundary couplings \reef{fbaction}. 

When  fields depend only on time, \ie dimensional reduction on $T^d$ where all circles are along the spacial coordinates,  the boundary of time is spacelike, \ie $n^\mu=(1,0,\cdots, 0)$ and $n^\mu n_\mu=-1$. When fields depend only on a spacial coordinate $x$, \ie dimensional reduction on $T^d$ where one of the circle is time direction, the boundary of $x$ is timelike, \ie $n^\mu=(0,-1,0,\cdots, 0)$ and $n^\mu n_\mu=1$. In both cases, using the gauge symmetries it is possible to write the metric, $B$-field  and dilaton as
 \beqa
G_{\mu\nu}=\left(\matrix{\mp n^2(\z)& 0&\cr 0&G_{ij}(\z)&}\right),\, B_{\mu\nu}= \left(\matrix{0&0\cr0&B_{ij}(\z)&}\right),\,  2\Phi=\phi+\frac{1}{2}\log|\det(G_{ij})|\labell{creduce}\eeqa
where the minus (plus) sign is for the case that $\z=t$ ($\z=x$). The laps function $n(\z)$ can also be fixed to $n=1$ ($n=-1$) for $\z=t$ ($\z=x$). The  reduction of the bulk action in \reef{baction} then becomes
\beqa
\bS_0^c&=&-\frac{2}{\kappa^2 n}\int d\z e^{-\phi}\Bigg[\frac{1}{4}\dB_{ij}\dB^{ij}-\frac{3}{4}\dG_{ij}\dG^{ij}-G^{ij}\dG_{ij}\dP-\dP^2+G^{ij}\ddot{G}_{ij}\Bigg]
\eeqa
where $\dG^{ij}\equiv G^{ik}G^{il}\dG_{kl}$ and dot refers to $\z$-derivative. Up to a total derivative term the above action can be written in $O(d,d)$-invariant form. In fact, using the following total derivative term:
\beqa
\int d\z \frac{d}{d\z}\Bigg[e^{-\phi}G^{ij}\dG_{ij}\Bigg]=\int d\z e^{-\phi}\Bigg[-G^{ij}\dG_{ij}\dP-\dG^{ij}\dG_{ij}+G^{ij}\ddot{G}_{ij}\Bigg]
\eeqa
one can write $\bS_0^c$ as
\beqa
\bS_0^c&=&-\frac{2}{\kappa^2n}\int d\z e^{-\phi}\Bigg[\frac{1}{4}\dB_{ij}\dB^{ij}+\frac{1}{4}\dG_{ij}\dG^{ij}-\dP^2\Bigg]-\frac{2}{\kappa^2n}\int d\z \frac{d}{d\z}\Bigg[e^{-\phi}G^{ij}\dG_{ij}\Bigg]
\eeqa
Since there is boundary, the total derivative term can not be ignored. It  can be transferred to the boundary by using the Stokes's theorem.

On the other hand, the  boundary is specified by $x^i=\sigma^i$ where $x^i$ does not include the coordinate $\z$. This coordinate  is independent of $\sigma^i$. Hence, the  reduction of $\sqrt{| g|}  e^{-2\Phi}=e^{-\phi}$. The unit vector to the boundary is fixed, \ie $\dot{n}=0$, and the cosmological reduction of the trace of the  extrinsic curvature becomes
\beqa
K^c=\frac{1}{2n}G^{ij}\dG_{ij}\labell{cK}
\eeqa
where $n=1$ ($n=-1$) for spacelike (timelike) boundary. Therefore, the reduction of the boundary term in \reef{baction} is ${\it exactly}$ cancelled with the total derivative term in the bulk action, \ie
\beqa
\bS_0^c&=&-\frac{2}{\kappa^2n}\int d\z e^{-\phi}\Bigg[\frac{1}{4}\Tr W^2-\frac{1}{4}\Tr Y^2-\dP^2\Bigg]\labell{S0c}\\
\prt\!\!\bS_0^c&=&0\nn
\eeqa
where $W=G^{-1}\dot{G}$, $Y=G^{-1}\dot{B}$.
The bulk action is invariant under $O(d,d)$ \cite{Veneziano:1991ek,Meissner:1991zj}.  Note that  the reduction of the  extrinsic curvature \reef{cK} can not be written in $O(d,d)$ invariant form. So it was necessary that this term was cancelled with the total derivative term in the bulk action. In other words, there is no way to write the boundary action in $O(d,d)$ invariant form unless it is zero.

At the higher order of $\alpha'$ the bulk and boundary actions should be invariant under $O(d,d)$. We will see  the couplings that are found by the $Z_2$-symmetry satisfy the $O(d,d)$ symmetry with no further constraint on the boundary parameters. However,  since the cosmological reduction of the leading order boundary action is zero we speculate that   the boundary action at all higher orders of $\alpha'$ to be zero  in a specific scheme in which the cosmological/one-dimensional  bulk action contains only first derivatives, as in the above leading order action, \ie \reef{cosm}. This can be a consistency check, and a constraint for the boundary couplings at higher orders of $\alpha'$, \eg the cosmological/one-dimensional  reduction of the boundary couplings \reef{fbaction} should satisfy the $O(d,d)$ symmetry and then the constraint \reef{cosm} fixes the coefficients of the $O(d,d)$ invariant terms to be zero.

Using a one-dimensional field-redefinition for which the lapse function remains invariant\footnote{Note that the lapse function $n(\z)$ in the cosmological reduction corresponds to the base space metric $\bg_{ab}$ in the circular reduction. The invariance of the lapse function is then consistent with the invariance of the base space metric  under the $Z_2$-transformations that we have considered in this paper.},  it has been shown in \cite{Meissner:1996sa} that the cosmological reduction of  the  bulk action \reef{fL1} can be written explicitly in  $O(d,d)$-invariant form when there is no boundary, \ie
\beqa
\bS_1^c&=&-\frac{2}{\kappa^2n}24a_1\int d\z e^{-\phi}\Bigg[\frac{1}{8}\Tr(Y^4)+\frac{1}{8}\Tr(W^4)+\frac{1}{4}\Tr(YWYW)-\frac{1}{2}\Tr(Y^2W^2)\nn\\&&-\frac{1}{16}[\Tr(Y^2)-\Tr(W^2)]^2-\frac{1}{2}[\Tr(Y^2)-\Tr(W^2)]\dot{\phi}^2-\frac{1}{3}\dot{\phi}^4\Bigg]\labell{coB}
\eeqa
  In this calculation   the terms which could not be written in terms of $O(d,d)$-invariant form are removable by field redefinition and total derivative terms. The total derivative terms which are needed to write the cosmological reduction of  the bulk couplings in the above  $O(d,d)$-invariant form,  are the following:
\beqa
&&\frac{24a_1}{n}\int d\z \frac{d}{d\z}\Bigg[e^{-\phi}\Tr(WY^2)+\frac{2}{3}e^{-\phi}\dot{\phi}^3+e^{-\phi}\dot{\phi}^2\Tr(W)+\frac{1}{2}e^{-\phi}\dot{\phi}(\Tr W)^2\nn\\&&\qquad\qquad-\frac{1}{2}e^{-\phi}\Tr(W)^2\Tr(W)+\frac{1}{4}e^{-\phi}(\Tr W)^3\Bigg]\labell{S1c}
\eeqa
  These total derivative terms have been ignored in \cite{Meissner:1996sa}  because it has been  assumed that the  spacetime has no boundary. In the presence of boundary, the above total derivative terms produce the following  boundary terms:
 \beqa
&&\frac{24a_1}{n}e^{-\phi}\Bigg[\Tr(WY^2)+\frac{2}{3}\dot{\phi}^3+\dot{\phi}^2\Tr(W)+\frac{1}{2}\dot{\phi}(\Tr W)^2\nn\\&&\qquad\qquad-\frac{1}{2}\Tr(W)^2\Tr(W)+\frac{1}{4}(\Tr W)^3\Bigg]\labell{tot}
 \eeqa
 which should be taken into account when studying   the $O(d,d)$-symmetry of the  boundary action \reef{fbaction}. For the timelike boundary  in which we are interested, $n=-1$.

 On the other hand, using the reductions \reef{creduce}, one  finds the  reduction of the timelike boundary action \reef{fbaction} for the following relation between the boundary parameters $b_1,b_{11}$:
 \beqa
 b_{11}&=&-24b_1\labell{b111}
 \eeqa
  to be
\beqa
&&24a_1e^{-\phi}\Bigg[-\Tr(WY^2)-\frac{2}{3}\dot{\phi}^3-\dot{\phi}^2\Tr(W)-\frac{1}{2}\dot{\phi}(\Tr W)^2+\frac{1}{2}\Tr(W)^2\Tr(W)-\frac{1}{4}(\Tr W)^3\Bigg]\nn\\&&+e^{-\phi}\Bigg[(24a_1+3b_1+\frac{b_{17}}{8})\dot{\phi}(\Tr W^2-\Tr Y^2)+(32a_1+4b_1-\frac{b_{18}}{6})\dot{\phi}^3\Bigg]
\eeqa
Note that the  reduction of the coupling in \reef{fbaction} with coefficient $b_{38}$ is zero. While the terms in the second line above are invariant under the $O(d,d)$ transformations, the terms in the first line are not. However, including  the residual total derivative terms from the bulk action, \ie \reef{tot}, one finds the boundary terms in the first line above are cancelled. Note that if one changes the coefficient of the boundary coupling $K_{\alpha }{}^{\gamma } K^{\alpha \beta } K_{\beta
\gamma } $ in \reef{fbaction}, then there would be the  term $\Tr( W)^3$ in the first bracket above which is not cancelled with the total derivative terms and is not invariant under $O(d,d)$ transformation. It means the cosmological reduction of the Euler character is not consistent with the $O(d,d)$ symmetry.

Therefore, up to a field redefinition, the  reduction of the bulk and boundary couplings for the relation \reef{b111} are given by the $O(d,d)$ invariant bulk action \reef{S1c} and the following $O(d,d)$ invariant boundary action:
\beqa
\prt \!\!\bS_1^c=-\frac{2}{\kappa^2} e^{-\phi}\Bigg[(24a_1+3b_1+\frac{b_{17}}{8})\dot{\phi}(\Tr W^2-\Tr Y^2)-(32a_1+4b_1-\frac{b_{18}}{6})\dot{\phi}^3\Bigg]\labell{bss}
\eeqa
Note that  the cosmological/one-dimensional  boundary action at the leading order is zero, \ie \reef{S0c}, hence, the field redefinition has no effect on the above  boundary action. 
Requiring the constraint \reef{cosm}, one finds the following two  relations between the parameters.
\beqa
b_{17}=-192 a_1-24b_1,\,b_{18}=192 a_1+24b_1
\eeqa
Inserting   the above relations  into the boundary action \reef{L12}, one  finally finds the timelike boundary action to be
\beqa
\prt \cL_1&\!\!\!\!=\!\!\!\!&b_{1}^{} \Bigg[-24K_{\alpha }{}^{\gamma } K^{\alpha \beta }
K_{\beta \gamma }+ 24 K^{\alpha }{}_{\alpha } K_{\beta \gamma
} K^{\beta \gamma } +24  K^{\alpha \beta } R_{\alpha
\beta } -24 K^{\gamma \delta } n^{\alpha } n^{\beta }
R_{\alpha \gamma \beta \delta }\nn\\&& +2
H^{\beta \gamma \delta } n^{\alpha } \nabla_{\alpha }H_{\beta
\gamma \delta } -48 K_{\beta \gamma } K^{\beta \gamma }
n^{\alpha } \nabla_{\alpha }\Phi  -6 H_{\alpha
}{}^{\delta \epsilon } n^{\alpha } n^{\beta } n^{\gamma }
\nabla_{\gamma }H_{\beta \delta \epsilon }\nn\\&& +H_{\beta \gamma \delta } H^{\beta \gamma \delta } K^{\alpha
}{}_{\alpha } + 12 K^{\alpha }{}_{\alpha } K_{\beta \gamma }
K^{\beta \gamma } - 12 K^{\alpha }{}_{\alpha } K^{\beta }{}_{
\beta } K^{\gamma }{}_{\gamma } - 3 H_{\alpha }{}^{\delta
\epsilon } H_{\beta \delta \epsilon } K^{\gamma }{}_{\gamma }
n^{\alpha } n^{\beta }\nn\\&& + 24 K^{\gamma }{}_{\gamma }
n^{\alpha } n^{\beta } R_{\alpha \beta } - 12
K^{\alpha }{}_{\alpha } R - 2 H_{\beta \gamma \delta }
H^{\beta \gamma \delta } n^{\alpha } \nabla_{\alpha }\Phi +
24 K_{\beta \gamma } K^{\beta \gamma } n^{\alpha }
\nabla_{\alpha }\Phi \nn\\&&+ 24 K^{\beta }{}_{\beta } K^{\gamma
}{}_{\gamma } n^{\alpha } \nabla_{\alpha }\Phi + 6 H_{\beta
}{}^{\delta \epsilon } H_{\gamma \delta \epsilon } n^{\alpha }
n^{\beta } n^{\gamma } \nabla_{\alpha }\Phi + 24 n^{\alpha }
R \nabla_{\alpha }\Phi - 48 K^{\beta }{}_{\beta }
\nabla_{\alpha }\Phi \nabla^{\alpha }\Phi \nn\\&&+ 48 K^{\gamma
}{}_{\gamma } n^{\alpha } n^{\beta } \nabla_{\alpha }\Phi
\nabla_{\beta }\Phi - 48 n^{\alpha } R_{\alpha
\beta } \nabla^{\beta }\Phi + 96 n^{\alpha } \nabla_{\alpha }
\Phi \nabla_{\beta }\Phi \nabla^{\beta }\Phi \nn\\&&- 96 n^{\alpha
} \nabla_{\beta }\nabla_{\alpha }\Phi \nabla^{\beta }\Phi -
96 n^{\alpha } n^{\beta } n^{\gamma } \nabla_{\alpha }\Phi
\nabla_{\beta }\Phi \nabla_{\gamma }\Phi + 96 n^{\alpha }
n^{\beta } n^{\gamma } \nabla_{\alpha }\Phi \nabla_{\gamma }
\nabla_{\beta }\Phi\Bigg]\nn\\&& + a_{1}^{} \Bigg[24 H_{\alpha }{}^{\gamma
\delta } H_{\beta \gamma \delta } K^{\alpha \beta } + 96
K^{\alpha }{}_{\alpha } K_{\beta \gamma } K^{\beta \gamma }
- 32 K^{\alpha }{}_{\alpha } K^{\beta }{}_{\beta } K^{\gamma
}{}_{\gamma } + 16 H^{\beta \gamma \delta } n^{\alpha }
\nabla_{\alpha }H_{\beta \gamma \delta }\nn\\&& - 192 K_{\beta
\gamma } K^{\beta \gamma } n^{\alpha } \nabla_{\alpha }\Phi
+ 192 K^{\beta }{}_{\beta } K^{\gamma }{}_{\gamma } n^{\alpha
} \nabla_{\alpha }\Phi + 192 n^{\alpha } R
\nabla_{\alpha }\Phi - 384 n^{\alpha } R_{\alpha
\beta } \nabla^{\beta }\Phi\nn\\&& - 384 K_{\alpha \beta }
\nabla^{\alpha }\Phi \nabla^{\beta }\Phi + 384 n^{\alpha }
\nabla_{\alpha }\Phi \nabla_{\beta }\Phi \nabla^{\beta }\Phi
- 768 n^{\alpha } \nabla_{\beta }\nabla_{\alpha }\Phi
\nabla^{\beta }\Phi \nn\\&&- 48 H_{\alpha }{}^{\delta \epsilon }
n^{\alpha } n^{\beta } n^{\gamma } \nabla_{\gamma }H_{\beta
\delta \epsilon } - 512 n^{\alpha } n^{\beta } n^{\gamma }
\nabla_{\alpha }\Phi \nabla_{\beta }\Phi \nabla_{\gamma
}\Phi + 768 n^{\alpha } n^{\beta } n^{\gamma }
\nabla_{\alpha }\Phi \nabla_{\gamma }\nabla_{\beta }\Phi\Bigg]\nn\\&& +
b_{38}^{} n^{\alpha } n^{\beta } n^{\gamma } n^{\delta }
\nabla_{\delta }\nabla_{\gamma }K_{\alpha \beta }\labell{final}
\eeqa
Then the effective actions are fixed up to one bulk parameter $a_1$ and two boundary parameters $b_1,\, b_{38}$.
 \section{Discussion}

 In this paper, we propose that the classical effective action of the string theory  at order $\alpha'^n$ in the presence of boundary, should satisfy the following three constraints:

 1-The effective action should be  a combination of the gauge invariant couplings that their coefficients should be independent of the geometry of the background, up to the field redefinitions. When  the background has a circle which is independent of its  boundary, then the dimensional reduction of the action  should satisfy the $O(1,1)$ symmetry.

 2-The effective action  should satisfy the least action principle with the boundary conditions that the massless fields and their derivatives up to order $n$ are arbitrary on the boundary. This boundary condition is consistent with the $O(1,1)$ symmetry.

 3-The cosmological/one-dimensional  reduction of the effective actions in a specific scheme in which only the first derivative terms appear in the bulk action should satisfy the $O(d,d)$ symmetry with zero boundary action, as in the leading order effective action.

Using the above constraints on the effective action  at order $\alpha'$, we have found the  bulk action \reef{fL1} up to one bulk parameter $a_1$, and the timelike boundary action \reef{final} up to the bulk parameter $a_1$ and two boundary  parameters $b_1,\, b_{38}$.

When the B-field and dilaton are zero, the gravity couplings in the bulk action are exactly the gravity couplings in the Gauss-Bonnet gravity, whereas the gravity couplings on the boundary have more couplings than those in Chern-Simons gravity.  Using the identity \reef{iden}, one can match the coefficient of the couplings in \reef{fbaction} which have the same structure as those in Chern-Simons gravity $Q_2$ for the following relations:
\beqa
b_1=-8a_1,\, b_{38}=32 a_1\labell{FF}
 \eeqa
The first relation could also be found by the least action principle in section 4 provided that one would not  discard the boundary total derivative terms. The gravity couplings in this case then become
\beqa
\!\!\bS_1+\prt\!\!\bS_1|_{\Phi=B=0}=-\frac{48a_1}{\kappa^2}\Bigg[\int d^Dx\sqrt{-G} R^2_{\rm GB}+\int d^{D-1}x\sqrt{-g}(Q_2+\frac{4}{3}n^{\alpha } n^{\beta }
\nabla_{\gamma }\nabla^{\gamma }K_{\alpha \beta })\Bigg]
\eeqa
While the Euler character  $I_2$  is not consistent with the symmetries of the classical effective actions, \ie it does not satisfy the $O(1,1)$ symmetry when the geometry has one  circle, nor with $O(d,d)$ symmetry when the geometry has the tours  $T^d$, the above couplings are consistent with the $O(1,1)$ and $O(d,d)$ symmetries. 

The effective actions for the timelike boundary that we have found  for  the relations \reef{FF}  are the following:
\beqa
\bS_1&=&-\frac{48 a_1}{\kappa^2}\int_M d^D x\sqrt{-G} e^{-2\Phi}\Big[R_{GB}^2+\frac{1}{24} H_{\alpha }{}^{\delta \epsilon } H^{\alpha \beta
\gamma } H_{\beta \delta }{}^{\varepsilon } H_{\gamma \epsilon
\varepsilon } -  \frac{1}{8} H_{\alpha \beta }{}^{\delta }
H^{\alpha \beta \gamma } H_{\gamma }{}^{\epsilon \varepsilon }
H_{\delta \epsilon \varepsilon }\nn\\&&  + \frac{1}{144} H_{\alpha
\beta \gamma } H^{\alpha \beta \gamma } H_{\delta \epsilon
\varepsilon } H^{\delta \epsilon \varepsilon }+ H_{\alpha }{}^{
\gamma \delta } H_{\beta \gamma \delta } R^{\alpha
\beta } -  \frac{1}{6} H_{\alpha \beta \gamma } H^{\alpha \beta
\gamma } R  -
\frac{1}{2} H_{\alpha }{}^{\delta \epsilon } H^{\alpha \beta
\gamma } R_{\beta \gamma \delta \epsilon }\nn\\&& -
\frac{2}{3} H_{\beta \gamma \delta } H^{\beta \gamma \delta }
\nabla_{\alpha }\nabla^{\alpha }\Phi + \frac{2}{3} H_{\beta
\gamma \delta } H^{\beta \gamma \delta } \nabla_{\alpha }\Phi
\nabla^{\alpha }\Phi + 8 R \nabla_{\alpha }\Phi
\nabla^{\alpha }\Phi + 16 \nabla_{\alpha }\Phi \nabla^{\alpha
}\Phi \nabla_{\beta }\nabla^{\beta }\Phi \nn\\&&- 16
R_{\alpha \beta } \nabla^{\alpha }\Phi \nabla^{\beta
}\Phi - 16 \nabla_{\alpha }\Phi \nabla^{\alpha }\Phi \nabla_{
\beta }\Phi \nabla^{\beta }\Phi + 2 H_{\alpha }{}^{\gamma
\delta } H_{\beta \gamma \delta } \nabla^{\beta
}\nabla^{\alpha }\Phi \Big]\labell{fL11}
\eeqa
\beqa
\prt\!\!\bS_1&=&-\frac{48 a_1}{\kappa^2}\int_{\prt M} d^{D-1} \sigma\sqrt{-g} e^{-2\Phi}\Bigg[ Q_2+\frac{4}{3}n^{\alpha } n^{\beta }
\nabla_{\gamma }\nabla^{\gamma }K_{\alpha \beta }+
\frac{2}{3} H_{\beta \gamma \delta } H^{\beta \gamma \delta }
n^{\alpha } \nabla_{\alpha }\Phi\nn\\&& - 2 H_{\beta }{}^{\delta
\epsilon } H_{\gamma \delta \epsilon } n^{\alpha } n^{\beta }
n^{\gamma } \nabla_{\alpha }\Phi -\frac{1}{3} H_{\beta \gamma \delta } H^{\beta
\gamma \delta } K^{\alpha }{}_{\alpha } +  H_{\alpha
}{}^{\gamma \delta } H_{\beta \gamma \delta } K^{\alpha \beta
} +  H_{\alpha
}{}^{\delta \epsilon } H_{\beta \delta \epsilon } K^{\gamma
}{}_{\gamma } n^{\alpha } n^{\beta }\nn\\&&- 16 K^{\gamma
}{}_{\gamma } n^{\alpha } n^{\beta } \nabla_{\alpha }\Phi
\nabla_{\beta }\Phi  + 16 K^{\beta }{}_{\beta }
\nabla_{\alpha }\Phi \nabla^{\alpha }\Phi- 16 K_{\alpha \beta } \nabla^{\alpha
}\Phi \nabla^{\beta }\Phi \nn\\&&- 16 n^{\alpha } \nabla_{\alpha
}\Phi \nabla_{\beta }\Phi \nabla^{\beta }\Phi +
\frac{32}{3} n^{\alpha } n^{\beta } n^{\gamma }
\nabla_{\alpha }\Phi \nabla_{\beta }\Phi \nabla_{\gamma
}\Phi\Bigg]\labell{finalb}
\eeqa
where $R_{GB}^2$ is the Gauss-Bonnet bulk couplings and $Q_2$ is the Chern-Simons boundary couplings \reef{I4}.  The bulk couplings for $a_1=1/96$ is the effective action of the bosonic string theory which has been found in \cite{Meissner:1996sa}. 

We have imposed the relation \reef{a112} to have standard propagators for the B-field. This relation can be also found by the $O(d,d)$ symmetry. We have seen that the cosmological reduction of the bulk couplings  are invariant under the $O(d,d)$ transformation up to some  total derivative terms which are not invariant. These anomalous terms  are exactly cancelled with the anomalous terms in the cosmological reduction of the boundary couplings. If one does not use the constraint \reef{a112}, then the two set of anomalous terms would cancel each other only under the condition \reef{a112}. We have performed this calculation explicitly. 

In the cosmological study, we have used   the scheme that the cosmological action has the first derivative of dilaton, \ie \reef{coB}, and the  boundary action has no term with first derivative of dilaton, \ie \reef{cosm}.   On the other hand, it has been shown in \cite{Hohm:2015doa,Hohm:2019jgu} that if one uses the most general  one-dimensional field redefinitions and uses integration by part, then  the cosmological reduction of the bulk  action at order $\alpha'$ and higher can be written in a scheme in which  the bulk action has only the first derivative of the generalized metric $\cS$, \ie no coupling involves the first derivative of dilaton. Moreover, the trace of two $\dot{\cS}$ is also removable by the laps function field redefinitions.  In the presence of boundary, the total derivative terms appear in the boundary by using the Stokes's theorem.   Hence, if one uses the scheme in which the derivative of dilaton does not appear in the bulk action, then in that scheme the boundary action may have the first derivative of the dilaton, \ie the cosmological boundary action may not be zero in that scheme.

In imposing the  $O(1,1)$ symmetry, we have assumed the   unit normal vector of the boundary in the base space, $n^a$ and its length remain  invariant under the T-duality transformations. This forces use to work with the most general gauge invariant bulk action \reef{L1bulk} which has 20 parameters, \ie we did not use the higher derivative field redefinitions to work with the independent bulk couplings. If one
 uses the most general field redefinitions, then  the  bulk action in the minimal scheme  has only 8 independent couplings. The T-duality fixes these parameters up to an overall factor \cite{Garousi:2019wgz}, \ie
\beqa
 \bS_1&=&\frac{-2c_1 }{\kappa^2}\int_M d^{D}x\, e^{-2\Phi}\sqrt{-G}\Big(   R_{\alpha \beta \gamma \delta} R^{\alpha \beta \gamma \delta} -\frac{1}{2}H_{\alpha}{}^{\delta \epsilon} H^{\alpha \beta \gamma} R_{\beta  \gamma \delta\epsilon}\nn\\
&&\qquad\qquad\qquad\qquad\qquad\quad+\frac{1}{24}H_{\epsilon\delta \zeta}H^{\epsilon}{}_{\alpha}{}^{\beta}H^{\delta}{}_{\beta}{}^{\gamma}H^{\zeta}{}_{\gamma}{}^{\alpha}-\frac{1}{8}H_{\alpha \beta}{}^{\delta} H^{\alpha \beta \gamma} H_{\gamma}{}^{\epsilon \zeta} H_{\delta \epsilon \zeta}\Big)\labell{S1bf}
\eeqa
where $c_1$ is the overall factor. For $c_1=1/4$, the above action is the effective action of the bosonic string theory at order $\alpha'$ which has been found in \cite{Metsaev:1987zx} by the S-matrix method. The above action and the action \reef{fL11} are related into each other by a particular field redefinition \cite{Meissner:1996sa}. In the presence of boundary, however, one may not be able to use the most general field redefinitions because they change the values of the massless fields and their derivatives on the boundary which may not be consistent with the least action principle. It
would be interesting to find the appropriate field redefinitions in the presence of the boundary to find the corresponding independent gauge invariant couplings and then  impose the constraints that we have studied in this paper, to  find independent bulk and boundary couplings at order $\alpha'$.   It would be also interesting to extend the calculation in this paper to find the boundary couplings at order $\alpha'^2, \alpha'^3$. The corresponding bulk actions in the minimal scheme have been found in \cite{Garousi:2019mca,Garousi:2020gio,Garousi:2020lof}.




\end{document}